\documentclass[3p]{elsarticle}

\usepackage{amsmath}
\usepackage{natbib}
\usepackage{amssymb}
\usepackage{graphicx}
\usepackage{dcolumn}
\usepackage{bm}
\usepackage{mathrsfs}
\usepackage{color}
\usepackage{comment}
\usepackage{psfrag}
\usepackage{url}
\graphicspath{{./Figures/}}
\usepackage{booktabs}

\usepackage[normalem]{ulem}
\usepackage{color}
\newcommand{\rev}[1]{{\color{black}#1}}
\newcommand{\revs}[1]{{\color{black}#1}} 
\newcommand{\revf}[1]{{\color{black}#1}}

\newcommand{\revCom}[1]{\color{black}#1} 

\newcommand{\et}{et al.}
\newcommand{\var}{\text{var}}
\newcommand{\cov}{\text{cov}}
\newcommand{\BE}{\mathbb{E}}
\newcommand{\CR}{\mathcal{R}}

\newcommand{\fig}{Fig.}
\newcommand{\sect}{Section}
\newcommand{\rp}{{\rm p}}
\newcommand{\rey}{\mbox{Re}}

\newcommand{\eg}{\textit{e.g.\ }}
\newcommand{\ie}{\textit{i.e.\ }}

\newcommand{\beq}{\begin{eqnarray}}
\newcommand{\eeq}{\end{eqnarray}}

\begin{document}

\title{Assessment of uncertainties in hot-wire anemometry and oil-film interferometry measurements for wall-bounded turbulent flows} 


\author[uu]{Saleh Rezaeiravesh}
\ead{saleh.rezaeiravesh@it.uu.se}
\author[kth]{Ricardo Vinuesa}
\ead{rvinuesa@mech.kth.se}
\author[uu,foi]{Mattias Liefvendahl}
\ead{mattias.liefvendahl@foi.se}
\author[kth]{Philipp Schlatter}
\ead{pschlatt@mech.kth.se}
\address[uu]{Division of Scientific Computing, Uppsala University, Sweden}
\address[kth]{Linn\'e FLOW Centre, KTH Mechanics, SE-100 44 Stockholm, Sweden \\
             and Swedish e-Science Research Centre (SeRC), Stockholm, Sweden }
\address[foi]{Swedish Defence Research Agency (FOI), Sweden}

\begin{keyword}
    Uncertainty quantification \sep 
    Wall-bounded turbulence \sep 
    Hot-wire anemometry \sep 
    Oil-film interferometry
\end{keyword}

\date{\today}

\begin{abstract}
In this study, the sources of uncertainty of hot-wire anemometry (HWA) and oil-film interferometry (OFI) measurements are assessed. 
Both statistical and classical methods are used for the forward and inverse problems, so that the contributions to the overall uncertainty of the measured quantities can be evaluated. 
The correlations between the parameters are taken into account through the Bayesian inference with error-in-variable (EiV) model. 
In the forward problem, very small differences were found when using Monte Carlo (MC), Polynomial Chaos Expansion (PCE) and linear perturbation methods. 
In flow velocity measurements with HWA, the results indicate that the estimated uncertainty is lower 
when the correlations among parameters are considered, than when they are not taken into account. Moreover, global sensitivity analyses with Sobol indices showed that the HWA measurements are most sensitive to the wire voltage, and in the case of OFI the most sensitive factor is the calculation of fringe velocity. 
The relative errors in wall-shear stress, friction velocity and viscous length are $0.44\%$, $0.23\%$ and $0.22\%$, respectively. 
Note that these values are lower than the ones reported in other wall-bounded turbulence studies. 
\revCom{Note that in most studies of wall-bounded turbulence the correlations among parameters are not considered, and the uncertainties from the various parameters are directly added when determining the overall uncertainty of the measured quantity. In the present analysis we account for these correlations, which may lead to a lower overall uncertainty estimate due to error cancellation.}
Furthermore, our results also indicate that the crucial aspect when obtaining accurate inner-scaled velocity measurements is the wind-tunnel flow quality, which is more critical than the accuracy in wall-shear stress measurements.\footnote{\textcopyright 2018. This manuscript version is made available under the CC-BY-NC-ND 4.0 license}

\end{abstract}

\maketitle

\section{Introduction}

Turbulent flows are extremely complicated due to the wide range of temporal and spatial scales present in them, responsible for various energy transfer mechanisms. The case of wall-bounded turbulence is even more complex due to the fact that the presence of the wall introduces an inhomogeneity in the wall-normal direction, which significantly affects the size of the turbulent structures. As discussed by Jim\'enez \cite{jimennez_structures}, at a particular wall-normal distance the energy transfer is on average from the largest, energy-containing scales towards the smallest, dissipative ones. However, due to the presence of the wall, whether a particular turbulent structure can be considered large or small depends on its wall-normal distance, a fact that increases the complexity of these flows.

Experimental uncertainty is a relevant topic in wall-bounded turbulence, due to the fact that small measurement errors may lead to very different conclusions regarding the nature of turbulent boundary layers (TBLs), especially when data at intermediate Reynolds numbers, $\rey$, are extrapolated to high-$\rey$ conditions. Note that the separation between the largest and smallest scales increases with Reynolds number. An example showing the relevance of accurate and independent measurement techniques is the relatively recent debate regarding the functional form of the so-called overlap region in TBLs, stirred among other factors by the different accuracies of the datasets analysed by various research groups. See for instance Refs.\ \cite{george_castillo,osterlund_et_al,barenblatt_et_al,monkewitz_et_al,noise} for further details on this topic. In this particular example the main quantity under investigation was the inner-scaled mean velocity profile $U^{+}(y^{+})$, where $U$ is the streamwise mean velocity, $y$ is the wall-normal location, and the superscript `$+$' denotes inner scaling as described in detail in Section \ref{sec:measure}.

Two widely used measurement techniques to experimentally determine the inner-scaled mean velocity profile are hot-wire anemometry (HWA) for the velocity, and oil-film interferometry (OFI) for the wall-shear stress. 
\revf{It is exactly these two methods that we study in this paper with reference to their measurement uncertainty.}
\revs{The measured quantities from combined HWA-OFI experiments can be used to estimate} the von K\'arm\'an coefficient $\kappa$, a very important parameter in wall-bounded turbulence research which is the inverse of the slope of the logarithmic layer in the overlap region, assuming that this is the functional form of the latter. Experimental uncertainties have led to multiple interpretations of the measurements, a fact that is illustrated in the work by Zanoun et al. \cite{zanoun_et_al}, where the reported value of $\kappa$ is represented as a function of the year (over seven decades), with values ranging from $0.32$ to $0.46$. The value of $\kappa$ reported by the Superpipe team in Princeton \cite{zagarola,mckeon} has also suffered changes over the years, a fact that could be explained by the different Pitot-tube probes used in the various studies, combined with the uncertainty in probe location for very high-$\rey$ and pressurized pipe-flow measurements \cite{vinuesa_superpipe}. Note that in Ref. \cite{orlu_superpipe} a documentation of their changes in other turbulence quantities is also provided.
\revs{By employing a Bayesian statistical tool, Oliver and Moser \cite{oliver12} studied the impact of the uncertainties in the experimental data of the flow mean and wall shear velocities on the overlap layer model parameters, including $\kappa$. The uncertainties in the data were assumed to be random and have specific distributions with presumed magnitudes close to what is expected from high-quality experiments.}
 The conclusions \revs{of the mentioned works} are complemented with the studies by Vinuesa et al. \cite{noise} and Segalini et al. \cite{segalini_et_al}, in which the influence of the measurement uncertainty in the determination of $\kappa$ are systematically evaluated.

Other relevant studies are the assessment of temporal and spatial resolution issues in hot-wire-anemometry found in Refs. \cite{hutchins_temporal} and \cite{hutchins_spatial}, respectively, the influence of temperature fluctuations in hot wires \cite{orlu_temp} and the evaluation of resolution issues in particle-image-velocimetry (PIV) measurements of turbulence quantities from Ref. \cite{segalini_piv}. 
In this context, the need for measurement corrections due to the underlying imperfections of the probes has been analyzed in a number of studies over the years. Some of these studies include the early work on Pitot tubes by MacMillan \cite{macmillan}, together with the more recent assessments by McKeon et al. \cite{mckeon_corrections} and Bailey et al. \cite{bailey}; the work on hot-wire corrections by Monkewitz et al. \cite{monkewitz_hw}, Smits et al. \cite{smits_hw} and Segalini et al. \cite{segalini_hw}; and the work by Vinuesa and Nagib \cite{vinuesa_nagib}, focused on Pitot tube measurements and wall-position of hot-wire probes. Note that a very important factor when establishing a canonical boundary layer is the flow development, as reported by Chauhan et al. \cite{chauhan_et_al} and Sanmiguel Vila et al. \cite{sanmiguel_vila_et_al}. Moreover, other recent studies have documented a dependence of the value of $\kappa$ on flow geometry \cite{variations} and the streamwise pressure under which the TBL develops \cite{vinuesa_aiaa,bobke_jfm}. In any case, it can be stated that there is some consensus in the wall-bounded turbulence community regarding the validity of the logarithmic law \cite{monkewitz_et_al,marusic_log_law,alfredsson_et_al}, with values of $\kappa$ between 0.38 and around 0.40 (as already discussed by von Karman in 1934 \cite{von_karman}).

Given the potential impact of measurement uncertainties in the conclusions drawn for experiments in wall-bounded turbulence, the aim of the present work is to \revs{implement relevant tools provided within the field of uncertainty quantification (UQ), see \eg \cite{smith,gelman}, to analyse the uncertainties involved in the HWA-OFI measurements and} characterize the sensitivity of such measurements to various factors, in order to identify the ones with the highest impact on the overall uncertainty. To this end, we consider velocity measurements obtained by means of HWA, as well as wall shear stress measurements with OFI. Although in some experimental studies these aspects have been partly addressed \cite{bailey}, a thorough identification of the underlying uncertainties, as well as their detailed uncertainty propagation, is lacking in the wall-bounded turbulence literature. 
We start from the basic quantities measured in HWA and OFI experiments, and perform a characterization of the forward propagation of uncertainties \revf{(known as forward problem \cite{smith})} in order to assess the respective contribution of all of these parameters to the final quantities, namely the flow velocity and wall-shear stress. 
\revs{There are various approaches in the UQ framework to perform the forward problem, ranging from the classical perturbation method to sample-based ones. 
A key aspect of the present study is to apply these methods to different forward problems involved in the HWA-OFI measurements. 
In addition, applying different approaches to tackle the inverse problems comprised of estimation of the model parameters appearing in different stages of the HWA-OFI measurements given uncertain data, is of central focus. 
In this context, it is shown how the parameter estimation approaches constructed to reflect a more realistic picture of the error structure of the measured data may estimate different values for the parameter uncertainties than the widely-used classical methods.

Besides employing the techniques that are less frequently used by the community of the experimentalists, it is interesting to show how the mathematical and statistical approaches developed in the UQ theoretical framework can be adopted to study a practical problem. 
To achieve this goal, the present article is structured to be self-contained up to some extent, providing the essence of the methods employed in different stages and citing relevant references for the interested readers. 
}

This article is structured as follows: in Section \ref{sec:UQ} the uncertainty quantification techniques employed in the present study are described in detail; in Section \ref{sec:measure} a general overview of the HWA and OFI measurement techniques is provided; in Section \ref{sec:Illust_Ex} the previously described techniques are applied to a HWA and an OFI experimental dataset, and the results are discussed; finally, a summary of the work and the main conclusions are provided in Section \ref{summary_conclusions}.

\section{Overview of the uncertainty quantification techniques}\label{sec:UQ}
Based on their nature, the uncertainties and errors can be generally categorized into two groups: first, the aleatoric uncertainties, which are also known as random errors, and cannot be reduced or removed by improving models or experiments since they are naturally inherent to the problem. 
The second type are epistemic or systematic uncertainties, which are usually biased and exist due to imperfections and discrepancies in models or experiments. 
Contrary to the aleatoric errors, the systematic uncertainties are not naturally defined in the probabilistic framework, \revs{see \cite{smith,oberkampf,rossi} and the references therein}. 
More specifically, uncertainties in laboratory experiments may stem from different sources such as incomplete or limited observed data, limited accuracy of the measurement devices, human-related errors, and other uncontrollable unknown sources.
Besides these, there might be errors due to mathematical models and formulas employed to describe the physical phenomena and to obtain quantities within the process of the experiment. 
These uncertainties may originate from model errors or discrepancies, a fact that implies that the mathematical relation is incapable of describing the true physics. This is another form of possible bias errors.

Therefore, various error sources are part of any experiment and cannot be completely eliminated due to physical constraints, technical infeasibility, and overwhelming expenses. 
However, there are techniques within the general framework of uncertainty quantification aiming at providing a better understanding of the relative importance of various identifiable uncertainties.
For uncertainty analysis of an interconnected complex system consisting of several smaller intermediate models, the general strategy is to go down to the factors residing at the lowest level and account for their associated uncertainties. 
This can be done directly via observing experiments, reading device manuals, making theoretical arguments, and relying on the experts' knowledge.  
Definitely, a deeper analysis of the system layers accounting for practical obstacles will lead to a better understating of the various sources of uncertainty. 
In the next step, the estimated uncertainties of the low-level quantities are propagated into the quantities and parameters of the connected models. 
These secondary uncertainties can be estimated by implementing different forward and inverse UQ techniques which are schematically illustrated in \fig ~\ref{fig:uqFrame}. 
A description of the various UQ concepts under consideration in the present work, is given below.

\begin{figure}[!t]
\centering
\includegraphics[scale=0.2]{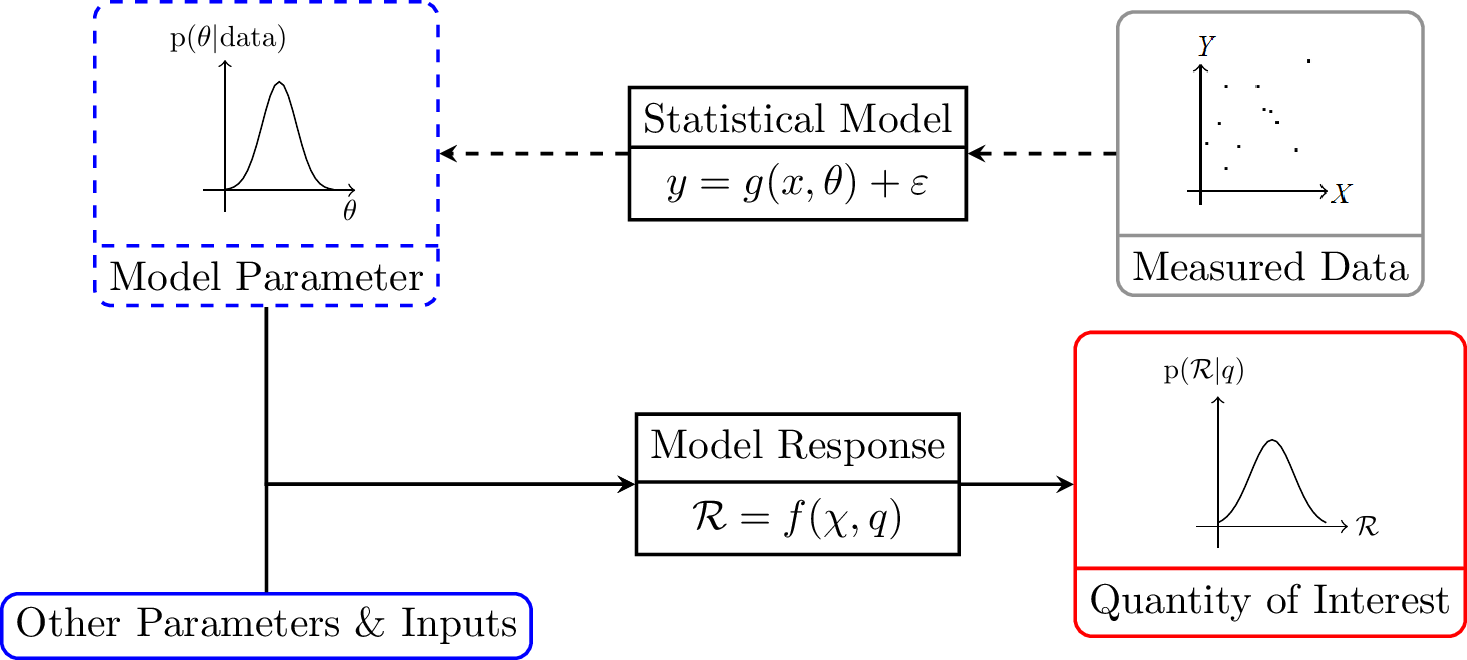}
\caption{Schematic illustrating the features provided within the UQ framework, including the inverse problem (dashed arrows) and the forward problem (solid arrows).
\revs{The grey box shows data observed for $(x,y)$ in an experiment, which are then employed to estimate parameters $\theta$ of model $y=g(x,\theta)$ within a statistical framework. These parameters along with other uncertain factors, all denoted by $q$ and shown by the blue boxes, propagate into response $\mathcal{R}$ through $\mathcal{R}=f(\chi,q)$, in which $\chi$ represents independent variables.} }\label{fig:uqFrame}
\end{figure}

\subsection{Forward problem}\label{sec:fwd}
\revs{As the starting point to explain the theoretical framework, consider the multi-variate response of a specific process, $\mathcal{R}\in \mathbb{R}^N$, can be  parametrised by the mathematical model}\revCom{:}
\begin{equation}\label{eq:uqMathMod}
\mathcal{R}=f(\chi,q)  \,,
\end{equation}
where $f$ is the model function, $\chi$ represents independent variables, and $q=(q_1,q_2,\cdots,q_p)$ denotes the model \revs{uncertain} parameter vector.
The aim of the UQ forward problem, also known as uncertainty propagation, is to propagate parameter uncertainties through model (\ref{eq:uqMathMod}) to construct \revs{the probability density function (PDF) of the model response or the quantities of interest (QoI), see \eg \cite{smith}}. 
\revs{The existence of these distributions reflects} the insufficient and incomplete knowledge about the QoI.
There are several methods to conduct the UQ forward problem\revs{, see \cite{smith,grabe,montomoli}.} \revs{Those that are used in the present paper} are briefly reviewed in the following.

The perturbation method is based on linearisation of the model $f(\chi,q)$ with respect to parameters $q$. 
This method for linearly parameterised models results in analytical expressions for QoIs, although it may not be a precise approach for highly non-linear models. 
If the parameters in (\ref{eq:uqMathMod}) are perturbed by $\delta q$ about a nominal state $\bar{q}$, then as \revs{for instance} shown in Ref. \cite{smith}, the mean and \revs{covariance matrix} of the response are given by: 
\begin{eqnarray}
\BE[f(\chi,q)] &=& f(\chi,\bar{q}) \, ,  \nonumber  \\ \label{eq:varPerturb}
\var[f(\chi,q)] &=& S^T V S \, . 
\end{eqnarray}
Here, \revs{$\BE$ denotes expected value}, $S$ is a matrix with local sensitivities $s_{ij}=\partial f_i/\partial q_j$ as elements, $f_i=f_i(q)=f(\chi_i,q)$, \revs{for $i=1,2,\cdots,N$}, and $V$ denotes the covariance matrix \revs{of parameters $q$}.
This approach was used by Bailey \et~\cite{bailey} to study uncertainties in zero-pressure-gradient TBL measurements using Pitot tubes, and by H{\"o}sgen \et~\cite{hosgen} to assess uncertainties in hot-wire measurements.

Linearisation can be avoided by employing sampled-based methods, in which PDFs of the QoIs are constructed by evaluation of the model response at sufficiently large numbers of realizations $q^{(j)}$, randomly sampled from the joint \revs{PDF} of the parameters, $\rho_Q(q)$, which is known either from a prior analysis of the measurements or by employing Bayesian techniques.
For sufficiently high number of samples, $n$, obtained by a Monte Carlo (MC) sampling technique, the \revs{estimates for the} mean and variance of the QoI are given by:
\begin{eqnarray*}
 \BE[f(\chi,q)] &=& \frac{1}{n}\sum_{j=1}^n f(\chi,q^{(j)}) \,,  \\
 \var[f(\chi,q)] &=& \frac{1}{n-1} \sum_{j=1}^n  \left[f(\chi,q^{(j)}) - \BE(f(\chi,q))\right]^2 \, . 
\end{eqnarray*}
The efficiency of this method is independent of the number of parameters, but is highly dependent on the representation of \revs{$\rho_Q(q)$}, \cite{smith}. 
The overall computational cost of the method relies on the MC sampling, for which the convergence rate is as slow as $\mathcal{O}(n^{-1/2})$. 
To improve this, other sampling algorithms such as the Latin Hypercube Sampling (LHS) method \cite{mckay}, with a convergence rate of $\mathcal{O}(n^{-1})$, can alternatively be used.

In order to avoid the computational cost of the sample-based methods, yet taking advantage of their ability to handle \revs{non-linear} models, spectral expansions can be used to produce \revf{an} approximate \revs{representation of $f(\chi,q)$}. 
\revs{In particular}, polynomial chaos expansions (PCE) \cite{ghanem:91,xiu_gPCE}, facilitate the construction of the statistical moments of the model response by decomposing \revs{$f(\chi,q)$} into separate deterministic and random components. 
A finite-dimensional representation of $f(\chi,q)$ as a function of mutually independent random variables $q \in \Omega \subset \mathbb{R}^p$, where $\Omega=\prod_{i=1}^p \Omega_i$, can be obtained as:
\begin{equation}\label{eq:pceExpans}
f(\chi,q) = \sum_{|{\bf k}|=0}^{K} \hat{f}_{{\bf k}}(\chi) \Psi_{{\bf k}}(q) \,,
\end{equation}
where $\hat{f}_{{\bf k}}(\chi)$ are deterministic coefficients of orthogonal polynomials $\Psi_{{\bf k}}(q)$, ${\bf k}=(k_1,k_2,\cdots,k_p)$ is a multi-index with magnitude $|{\bf k}|=\sum_{i=1}^p k_i$, \revs{and $K$ is \revf{a} positive integer}. 
The $p$-variate basis functions are defined as \revs{${\Psi_{{\bf k}}(q)=\prod_{i=1}^p \psi_{k_i}(q_i)}$}.
The random nature of the process is expressed in terms of the polynomial chaos bases $\psi_{k_i}(q_i)$, which are chosen from well-known \revs{sets of orthogonal polynomials} depending on the type of distribution of the random parameters $q_i$\revs{, see Ref. \cite{xiu_gPCE}}.
\revs{By construction}, the polynomials form a complete orthogonal set of basis functions to span the random parameter space, \revs{which means that} the deterministic coefficients are determined from:
$$
\hat{f}_{k}(\chi) = \frac{1}{\xi_{k}} \BE[f(\chi ,q) \Psi_{k}(q)] \ ,
$$
where $k$ is the single-index obtained by reordering the multi-indices and $\xi_k=\BE [\Psi_k(q)\Psi_k(q)]$.  
Eventually, the mean and variance of the model response is obtained in terms of the deterministic coefficients and polynomials (see Refs. \cite{ghanem:91,xiu}) as:
\begin{eqnarray}
\BE[f(\chi,q)]&=&\hat{f}_{0}(\chi) \ , \label{eq:pceMean} \\\label{eq:pceVar}
\var[f(\chi,q)]&=&\sum_{k=1}^N \hat{f}_{k}^2(\chi) \xi_k \,.
\end{eqnarray}
For numerical evaluation of the integrals involved in obtaining the deterministic coefficients, different deterministic and stochastic quadrature techniques may be employed. 
In the case where the parameters are not mutually independent, it is required to transform them into a set of independent random variables to which the above formulation can be applied.  
The reader is referred to Refs. \cite{ghanem:91,xiu} for the details of these techniques, as well as for further discussion regarding the PCE method.

\subsection{Sensitivity analysis}
In connection with the uncertainty propagation problem and as a complement to it, a sensitivity analysis can be performed to determine how variations in the model parameters affect the model response. 
As a result of this analysis, the most influential parameters are distinguished from those which are comparatively less significant. 
In the most widely used approach, known as local sensitivity analysis (LSA), only one model parameter at a time is perturbed around a nominal state, which can be its mean value, and the associated variation of the model response is \revs{evaluated}. 
This is directly connected to the perturbation method described in the previous section. 
In fact, the local sensitivity indices are the \revf{components} \revs{$s_{ij}$} of $S$ in equation (\ref{eq:varPerturb}).

In contrast, as a complement to the sampled-based or spectral expansion methods for the forward problem, a global sensitivity analysis (GSA) can be conducted. 
In addition to being independent of any linearisation and being capable of considering interactions between parameters, the main advantage of this analysis is its ability to evaluate the sensitivity of the QoI with respect to different uncertain parameters when these parameters change throughout their whole admissible spaces.    
\revCom{For the details of the technique used for GSA in the present study, see \ref{sec:GSA_appendix}.}

\subsection{Inverse problem}\label{sec:uqInv}
\revs{Consider the mathematical model:}
$$
y=g(x,\theta) \, ,
$$ 
for which a set of observed data is available for $y$ at corresponding $x$.
The aim of the inverse problem is to estimate the unknown parameters $\theta$ (note that as a convention, when the model parameters are unknown and to be estimated in the inverse problem, the symbol $\theta$ is used to denote a contrast with the known $q$ in the forward problem). 
To this end, \revs{techniques developed within two main frameworks} can be employed, \revs{see \eg Ref. \cite{tarant}}.
Within the so-called frequentist point of view, the model parameters are assumed to be fixed but unknown. 
Together with the associated confidence intervals, they are estimated for instance by the least-squares (LS) method, to minimize the summation of squares of the errors between the model and observations. 
In contrast, in the Bayesian \revs{perspective}, the unknown model parameters are considered to be random and treated in a subjective manner. 
This allows us to update our initial belief about a parameter as new experiment is performed.

To estimate the model parameters $\theta$ in a Bayesian framework, a statistical model corresponding to the above mathematical model is required, \revs{which for an additive error form is written as:} 
\begin{equation}\label{eq:statisModelInverse}
Y =G(\theta)+\epsilon \,.
\end{equation}
Here, $Y=\{y_i:i=1,2,\cdots, N\}$ denotes the observed data and $G=\{g_i:i=1,2,\cdots, N\}$, where $g_i(\theta)=g(x_i,\theta)$.
The error $\epsilon=\{\varepsilon_1,\varepsilon_2,\cdots,\varepsilon_N\}$ in its general form includes both model inadequacy (bias) and measurement errors. 
\revCom{
The structure of the error term relies on both the model function $g$ and the observed data, see Refs. \cite{kennedy,smith}. 
If $g$ is completely physics-based, then it can predict the true mean of $y$, given true means of the input $x$ and the parameters. 
In this case, the error term only accounts for the error in the observed data as well as the random residuals. 
In contrast, if $g$ is not physics-based, then its prediction may show some deviations (with recognizable patterns) from the true mean of $y$, even when true input and parameter values are used. 
In this case, one would have to include a model inadequacy (bias) term in the error $\varepsilon$ shown in \fig~\ref{fig:uqFrame}, to account for ``unaccommodated'' model discrepancies. 
}
\revCom{
For simplicity, in the present work we consider model responses $g(x,\theta)$ with no bias, so they are assumed to have negligible discrepancy. Since the bias is not considered in the analysis, $\varepsilon_{i}$ corresponds exclusively to random measurement errors.}
\revs{This is motivated by the fact that when calibrated parameters $\theta$ based on the sample data for $(x_i,y_i)$ with $i=1,2,\cdots,N$ are used, \revf{the} model $y=g(x,\theta)$ is capable of predicting the mean values $\bar{y}_i$ associated to $\bar{x}_i$, for $i=1,2,\cdots,N$, without the requirement of adding a bias term.}

\revs{As it is explained below, different interpretations and \revf{models}, ranging from simple to more involved, can be considered for $\epsilon$ to reflect the structure of the errors of the observed data, which as shown in Sections \ref{sec:hwaEx_factors} and \ref{sec:ofiEx_factors}, affect the uncertainty of the estimated parameters $\theta$.}

If our prior information of the parameters $\theta$ before conducting any experiment and gathering data is represented by prior density $\rp_0(\theta)$, then the corresponding posterior density, $\rp(\theta|Y)$, which specifies the distributions of parameters based on the sampled observations, is achieved from the \revs{Bayes' theorem of inverse problems} (see \eg Ref.~\cite{gelman}):  
\begin{equation}\label{eq:bayes}
\rp(\theta|Y)\propto \rp(Y|\theta) \rp_0(\theta) \, .
\end{equation}
This is the role of the likelihood density $\rp(Y|\theta)$, namely to transfer the information provided by the observations and hence update the posterior density of the parameters.

If the errors $\varepsilon_i$ are assumed to be independent and identically distributed (iid) with $\varepsilon_i \sim \mathcal{N}(0,\sigma)$, then $y_i \sim \mathcal{N}(g_i(\theta),\sigma)$, where $\theta$ and $\sigma$ are unknown true model and error parameters, respectively. 
According to this assumption, the likelihood function is:
\begin{equation}\label{eq:GaussLikeli}
L(\theta,\sigma^2|Y)=(2\pi \sigma^2)^{-N/2} \exp \left( -\frac{1}{2\sigma^2} \sum_{i=1}^N (g(x_i,\theta)-y_i)^2 \right) \,.
\end{equation}
In this formulation, \revs{$x$ represents the independent variables that are certain.} However, in the experiments we may encounter situations where both $x$ and $y$ are uncertain. 
By adopting additive error models for both of these variables within the classical errors-in-variable (EiV) form \cite{dellaportas}, we have:
\begin{equation}\label{eq:bayes_EiV}
	\begin{cases}
		Y=g(X_t,\theta) +\epsilon_y \\
		X=X_t +\epsilon_x \\
	\end{cases} \, ,
\end{equation}
where $X$ and $Y$ are uncertain observations for actual but unknown $X_t$, and $Y_t$, respectively, related by the errors $\epsilon_x$ and $\epsilon_y$.
The Bayes' formula for this case is written as:
\begin{equation}\label{eq:bayes_EiV_formul}
\rp(\theta,\theta_x,\theta_y|Y,X) \propto \rp(Y|\theta,\theta_y ,X,\theta_x,X_t) \rp_0(\theta,\theta_x,\theta_y,X_t) \, ,
\end{equation}
which provides a relation for the posterior densities of the model parameters $\theta$ and error parameters $\theta_x$ and $\theta_y$, given observed data for both $x$ and $y$. 
The implementation of this case, especially the construction of the likelihood function, is more involved compared to the first case, and is explained in detail in \ref{sec:EiV_likeli}, taking into account the specifications of the situations we deal with in the present study.

Throughout this work, when the Bayesian inference is conducted for the purpose of estimating the \revs{model} parameters, the associated prior densities are taken to be non-informative meaning that they are uniform over a range of parameter admissibility. 
In cases where the aim is to estimate the error variance $\sigma^2$ of Gaussian errors, the prior is taken to be in the inverse-gamma family in order to take advantage of the conjugacy property \cite{gelman}.
The prior distributions of unknown $X_t$ in the EiV model are assumed to be multivariate Gaussian, a choice for which an improved convergence of the Bayesian inference was \revs{obtained}.

\subsection{\revCom{Implementation of the UQ techniques}}
\revCom{
In the present study, the Dakota library \cite{dakotaMan} was used to produce the numerical results for the sample-based and spectral expansion methods in the forward problems, and also to obtain the Sobol sensitivity indices. For this purpose, an in-house computer code was developed based on the models described in \sect ~\ref{sec:measure}. This code is linked to Dakota through an interface  in a non-intrusive way.  
}

\revCom{
To obtain the posterior density in a Bayesian inverse problem, it is required to numerically evaluate a multi-dimensional integral corresponding to the Bayes' formula.
The relatively large dimension of this integral which is determined by the number of parameters including model, errors, and actual unobserved $X_t$ (when equation (\ref{eq:bayes_EiV_formul}) is used), necessitates employing an efficient algorithm for sampling from the posterior density. 
In particular, the Adaptive Multi-Level Algorithm \cite{prudencio12} as implemented in the QUESO library \cite{queso,queso2}, is used in the present study to produce the numerical results for the Bayesian inverse problems in \sect ~\ref{sec:Illust_Ex}. 
}

\section{Overview of the measurements}\label{sec:measure}

In the following section, in order to illustrate the UQ method on the example of predicting the mean-velocity profile in turbulent wall-bounded flows, the corresponding measurement techniques are introduced, together with a discussion of the various uncertainties and error sources.

\subsection{Hot-wire measurements of flow velocity}\label{sec:HWAmeasure}
\begin{figure}[t!]
\centering
\includegraphics[scale=0.73]{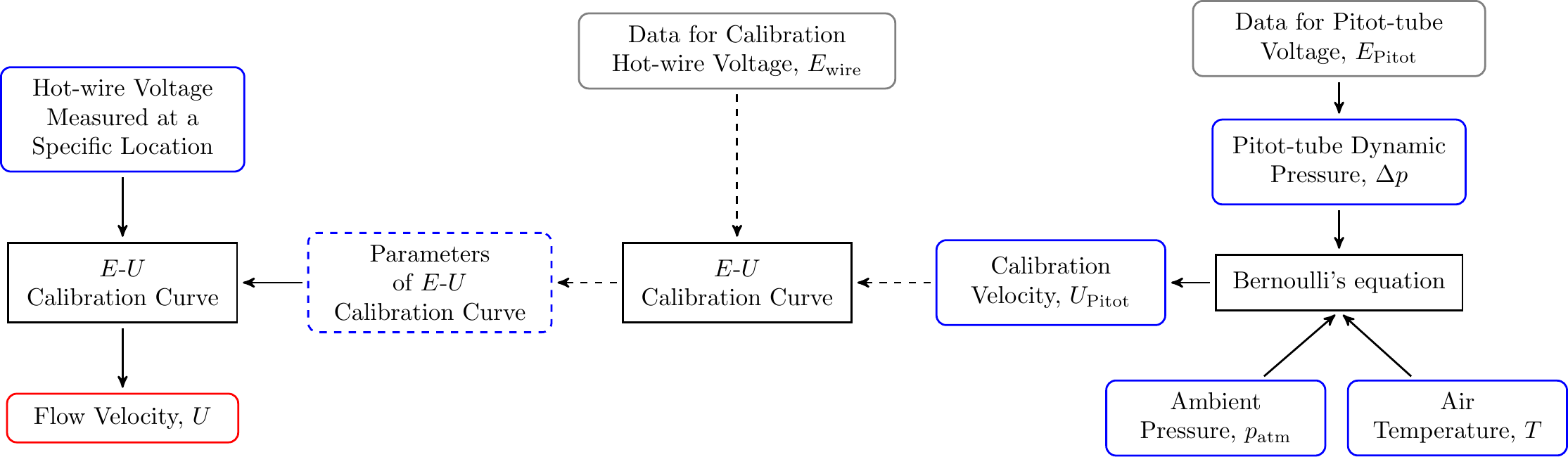}
\caption{Uncertainty propagation into the velocity measured by HWA. \revs{Same conventions as in \fig ~\ref{fig:uqFrame}.}}
\label{fig:hwaUQ}
\end{figure}

The first QoI under consideration is the flow velocity $U$, and in particular in the present study we will focus on the uncertainty from hot-wire anemometry measurements. As discussed in, for instance, Ref. \cite{orlu_vinuesa}, HWA is based on measuring the changes in forced convection experienced by a small heated wire as the flow velocity changes. This implies that HWA is an indirect measurement technique, and therefore relies on other measurements in order to obtain the experimental value of $U$. It is therefore interesting to assess the impact of the various measured variables on the final uncertainty of $U$. The idea is that the hot wire is electrically heated, and the flow velocity $U$ is the main factor for heat loss. If the HWA is operated at a constant overheat ratio \cite{orlu_vinuesa}, then the flow velocity $U$ can be related to the voltage drop across the wire $E$ as follows:
\begin{equation}\label{kings_law}
E^{2}=A+BU^{n},
\end{equation} 
where $A$, $B$ and $n$ are calibration constants. Equation (\ref{kings_law}) is known as King's law \cite{king}, although it is \revs{also} common to use higher-order equations to relate $E$ and $U$. In the so-called {\it in situ} calibration process, one places the hot-wire anemometer close to a Pitot-static tube, and the flow velocity is varied in order to produce a calibration curve relating the measured voltage $E$ with the flow velocity $U$. The Pitot-static tube measures the difference between the total and the static pressure, $\Delta p$, which corresponds to the dynamic pressure. From Bernoulli's equation, the velocity can be obtained from:
\begin{equation}\label{Ucal}
U=\sqrt{\frac{2 \Delta p}{\rho}},
\end{equation} 
where $\rho$ is the fluid density. Note that $\rho = p_{{\rm atm}} / ( R T )$, where $p_{{\rm atm}}$ is the ambient pressure, $T$ is the air temperature and $R$ is the gas constant which takes a value of 287 J/(kg K) in the case of air. A summary of the variables contributing to the uncertainty of the measured value of $U$ is shown in Fig. \ref{fig:hwaUQ}, together with the uncertainty propagation. As described above, the hot-wire calibration curve is obtained from the velocity measurements of a Pitot-static tube. The voltage from the Pitot reading $E_{\rm{Pitot}}$ is converted to a pressure reading through the specifications of the Pitot tube. This pressure reading is then transformed into a velocity through equation (\ref{Ucal}), together with readings of ambient pressure and temperature. The velocity reading from the Pitot is then used to produce the calibration curve (\ref{kings_law}), together with the hot-wire voltage reading. This calibration curve  then provides, for each wall-normal location, the velocity measured by the hot-wire anemometer in terms of the measured voltage $E_{\rm{wire}}$. Further details of the dataset under consideration in the present work are given in Section \ref{sec:Illust_Ex}.

\subsection{Oil-film interferometry measurements of wall-shear stress}\label{sec:OFImeasure}
The second QoI is the friction velocity $u_{\tau}=\sqrt{\tau_{w} / \rho}$, with $\tau_{w}$ being the wall-shear stress. The friction velocity is an important quantity in wall-bounded turbulence research, since it is used \revs{in the TBL inner scaling} to non-dimensionalize the velocity and fluctuation profiles. In fact, an accurate determination of $u_{\tau}$ is essential in order to properly interpret high-Reynolds number turbulent boundary layer measurements \cite{noise,vinuesa_nagib,vinuesa_superpipe}. One of the most widely used experimental methods of determining the wall-shear stress in wall-bounded turbulent flows is oil-film interferometry, which leads to accuracies in $\tau_{w}$ measurements on the order of $1\%$ \cite{channel_ofi}. OFI is based on the relation between the development of a thin oil film and the wall-shear of the incoming stream driving it (see for instance Refs. \cite{tanner_blows,fernholz_et_al,ruedi,ofi_segalini,vinuesa_orlu}). Experimentally, several oil drops are deposited on the wall, and the incoming stream drives the oil drop forming a thin film of height $h$, which is a function of the streamwise distance $x$ and of the time $t$. Tanner and Blows \cite{tanner_blows} proposed in 1976 a method to measure the film height as a function of $x$ and $t$, based on the formation of Fizeau interferometric fringes. These fringes can easily be observed when illuminating the film with a monochromatic source of light, where sodium lamps are the most widely used. As the film develops, the variation of $h$ leads to a change in the optical path length of the incoming light within the oil film. This produces an interference between the beam that is directly reflected from the oil-film surface and the one that travelled through the film, is reflected on the wall, and  refracted back outside of the film. This interference can be either constructive or destructive, leading to a fringe pattern which is used to characterize the downstream evolution of the oil film.   The thickness difference between two consecutive fringes $\Delta h$ can be computed as \cite{hccht}:
\begin{equation}\label{Dh}
\Delta h=\frac{\lambda}{2\sqrt{n_{{\rm oil}}^{2}-n_{{\rm air}}^{2} \sin^{2} \alpha }},
\end{equation}
where $\lambda$ is the wavelength of the illuminating light (in the case of sodium $\lambda=589.3$ nm), $n_{{\rm air}} \simeq 1$ and $n_{{\rm oil}}$ are the refractive indices of air and oil, respectively, and $\alpha$ is the angle between the camera and the wall-normal axis. The wall-shear stress depends on the value of $\Delta h$ obtained with equation (\ref{Dh}), and on the value of the oil kinematic viscosity $\nu$. It is therefore crucial to accurately calibrate the oil viscosity in order to obtain the dependence of $\nu$ with temperature, since any error in the actual value of $\nu$ directly affects the resulting $\tau_{w}$. The importance of the oil calibration procedure is discussed in Ref. \cite{ruedi}. Moreover, the value of $\tau_{w}$ also depends on the time evolution of the fringes, in particular of the time variation of the fringe spacing ${\rm d} \lambda_{f} / {\rm d} t$, which is computed by using the wavelength estimation method as described for instance in Ref. \cite{channel_ofi}. The wall-shear stress is then computed as:
\begin{equation}\label{eq:tauOFI}
\tau_w= \frac{\mu_{\rm oil}}{\Delta h} \frac{{\rm d} \lambda_f}{{\rm d} t},
\end{equation}
where $\Delta h$ is obtained from (\ref{Dh}). Additional details on the calculation procedure can be found in Ref. \cite{vinuesa_orlu}.

\begin{figure}[ht!]
\centering
\includegraphics[scale=0.9]{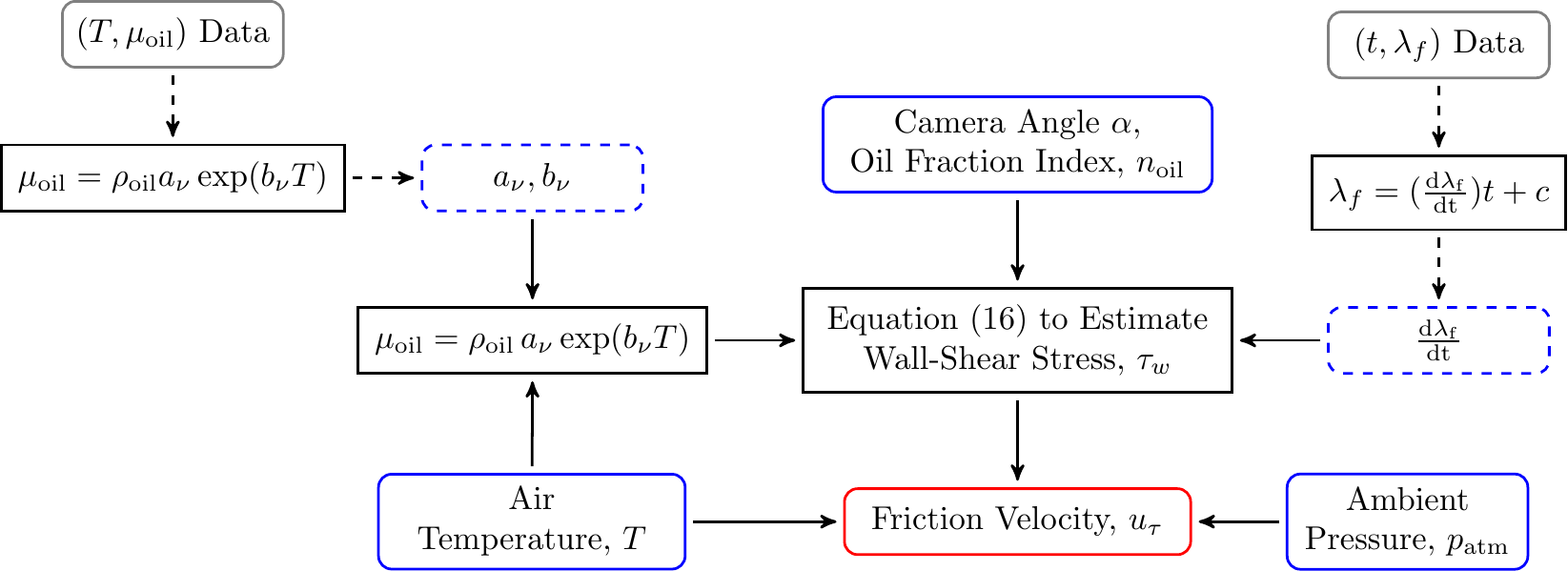}
\caption{Uncertainty propagation into the friction velocity obtained by OFI. \revs{Same conventions as in \fig ~\ref{fig:uqFrame}.} }\label{fig:ofiUQ}
\end{figure}

The uncertainty propagation diagram for the OFI measurements is shown in Fig. \ref{fig:ofiUQ}, where the main idea is to use equations (\ref{eq:tauOFI}) and (\ref{Dh}) to calculate the wall-shear stress $\tau_{w}$. To use these equations we require the time-variation of the spacing between fringes ${\rm d} \lambda_{f} / {\rm d} t$ (which is obtained through the global wavelength estimation method applied to the oil-film pictures showing the interferometric patterns), the oil viscosity (which is obtained through the calibration curve), as well as the camera angle and the oil refractive index. Ambient temperature and pressure are also required to compute the friction velocity from the wall-shear stress. The dataset considered in the current analysis is described in Section \ref{sec:Illust_Ex}.

\revCom{
As stated in \sect~\ref{sec:uqInv}, the present analysis is focused on random measurement errors, and the bias is not included. 
Therefore, effects such as the overestimation of $u_{\tau}$ obtained when using the theory by Tanner and Blows \cite{tanner_blows}, as suggested by Segalini et al. \cite{ofi_segalini}, are not accounted for. 
An adequate modelling of the bias in the statistical model is a challenging task, which will be included in future extensions of the present study.
}

\section{Uncertainties in HWA and OFI measurements}\label{sec:Illust_Ex}

In this section \revs{a complete procedure is described} \revs{for} how the UQ techniques \revs{reviewed} in \sect ~\ref{sec:UQ} can be employed to assess the uncertainties involved in different stages of HWA and OFI experiments as described in \sect ~\ref{sec:measure}. To this end, \revs{the uncertainties of the} datasets of the observations made for the quantities in the \revs{grey} boxes in \fig ~\ref{fig:hwaUQ} and \fig ~\ref{fig:ofiUQ} in laboratory experiments are first studied \revs{and then employed to estimate the different model parameters involved in HWA and OFI measurements} in Sections \ref{sec:hwaEx_factors} and \ref{sec:ofiEx_factors}, respectively. 
These \revs{estimated parameters along with other factors} are then used to investigate the propagation of the associated uncertainties in the direction of the \revs{solid} arrows \revs{in \fig ~\ref{fig:hwaUQ} and \ref{fig:ofiUQ} into the QoIs} in \sect ~\ref{sec:ofi_hwaFWD}. 
\revf{
It is emphasized that upon availability of information about the uncertainties of the quantities constituting the factors currently residing in the grey and solid-line blue boxes, the UQ analysis could be refined. 
}

In the analysis conducted here, a main basic assumption is that all errors \revf{contaminated} the \revs{experimentally-observed data} are of random type and hence no bias exists.
\revs{This, along with the assumption that all model responses \revf{whose parameters are} estimated have no discrepancy, implies that the error in \revf{equation} (\ref{eq:statisModelInverse}) is random. Hence, this error can be modelled through the different approaches reviewed in \sect ~\ref{sec:uqInv}.} 
\revCom{
Although we do not study the influence of the bias in the present analysis, one way to detect sources of bias would be to compare measurements of the same flow case from various facilities, performed by different researchers, as in the International Collaboration on Experimental Turbulence (ICET) described in Bailey \et~\cite{bailey}. 
There are essentially two types of bias errors: the ones associated to the mathematical model, and those present in the measurements. 
An example of the first type would be due to imperfect physics-based models, such as in the theory by Tanner and Blows \cite{tanner_blows}, as discussed by Segalini \et~\cite{ofi_segalini}. 
There are some examples of bias in the measurements, which can be corrected through physics-based or partially-physics-based models, as observed in the attenuation of velocity fluctuations in hot-wire anemometry, \cite{monkewitz_hw,segalini_hw,smits_hw}, or the corrections necessary for Pitot-tube measurements, \cite{bailey,mckeon_corrections,vinuesa_nagib}. 
Finally, some examples of bias present in the measurements which are not corrected are the effect of inadequate tripping of the boundary layer, \cite{sanmiguel_vila_et_al}, insufficient development length, \cite{chauhan_et_al}, or differences among researchers when performing OFI measurements or manufacturing hot-wire anemometers.
}

The \revs{datasets} under consideration for the uncertainty assessment of HWA consist of simultaneous hot-wire and Pitot-static tube measurements at a total of 11 freestream velocities ranging from 0 (no-flow condition) to 22 m/s. 
These measurements were taken at a single wall-normal location, in one of the \revf{open-return} wind tunnels at the Illinois Institute of Technology (IIT, Chicago) used for teaching purposes. 
\revf{In this tunnel, the square test section has a cross-sectional area of $381 \times 381$ mm, and is 420 mm long in the streamwise direction. A total of four honeycomb screens are used to condition the inflow before a 3:1 contraction, and turbulence intensity levels between $0.6$ and $1\%$ were measured in various streamwise locations.}
The reason to choose \revs{these datasets}, with lower accuracy than the ones observed in high-quality wall-bounded turbulence research, is to highlight the contributions from the various parameters contributing to the HWA uncertainty.

\revCom{
Some of the parameters characterizing the wind-tunnel flow quality are the temperature and pressure fluctuations, as well as the freestream turbulence intensity \cite{reshotko}. 
For instance, Nagib \et~\cite{nagib94} reported temperature fluctuations below $1\%$, static pressure fluctuations of around $0.4\%$ of the freestream dynamic pressure, and freestream turbulence intensities between $0.02\%$ and $0.05\%$ in the National Diagnostic Facility (NDF) at IIT. These values, which are considered to be adequate for wall-bounded turbulence research, should be complemented with an adequate flow development, as discussed by Chauhan \et~\cite{chauhan_et_al} or Sanmiguel Vila \et~\cite{sanmiguel_vila_et_al}.
}

Each of the datasets consists of $10^{4}$ samples for Pitot-tube and hot-wire voltages, and the main use of these measurements was to obtain calibration curves for the hot-wire anemometer. 
Note that before storing samples, the tunnel was run for sufficiently long times to allow the flow conditions to settle, and therefore the various datasets can be considered to be independent and thus uncorrelated.

The OFI dataset was obtained from the channel-flow measurements described in Ref. \cite{channel_ofi}. This dataset includes the oil calibration data, the camera angle and oil refractive index, together with a series of oil-fringe pictures corresponding to one of the measurements. These allowed to determine the analysis region producing the interferometric pattern required to obtain ${\rm d} \lambda_{f} / {\rm d} t$ based on the global wavelength estimation method.

\revs{
As it is frequently used in the following discussions, we adopt the following terminology: 
the percentage of the uncertainty in the physical quantity $\varphi$ can be calculated by: 
$
\zeta_\varphi= {\sigma_{\varphi}}/{\bar{\varphi}} \times 100
$, 
where $\sigma_{\varphi}$ and $\bar{\varphi}$ denote the associated standard deviation and mean values of $\varphi$, respectively. 
However, as it is conventional, see \eg Ref. \cite{bailey}, what is reported by the experimentalists is the uncertainty in $95\%$-confidence sense $\zeta_{\varphi,95\%}$, which is equal to $1.96\,\zeta_{\varphi}$. 
}

\subsection{Accounting for the uncertainties in HWA measurements, calibration phase}\label{sec:hwaEx_factors}
\begin{figure}[t]
\centering
\includegraphics[scale=0.53]{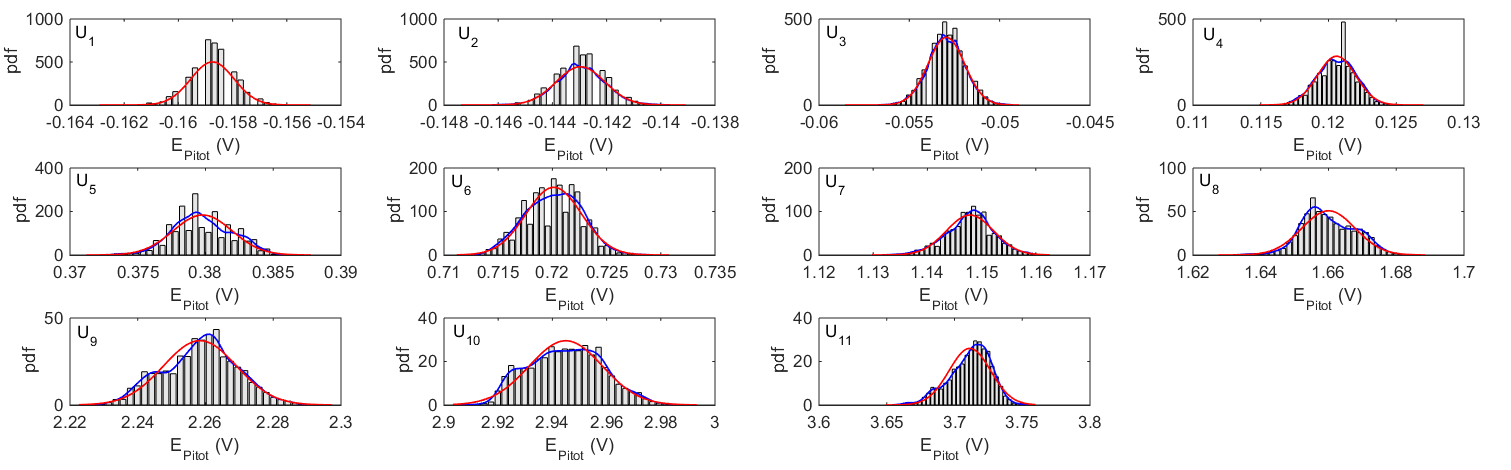}
\caption{Histograms and \revs{PDF (blue lines) of $10^4$ Pitot-tube voltage samples experimentally measured at 11 freestream velocities, denoted by $(U_1,U_2,\cdots,U_{11})$, along with the closest Gaussian distributions fitted to the samples (red lines).}
 }\label{fig:voltagePitotPdf}
\end{figure}

\begin{figure}[t]
\centering
\includegraphics[scale=0.53]{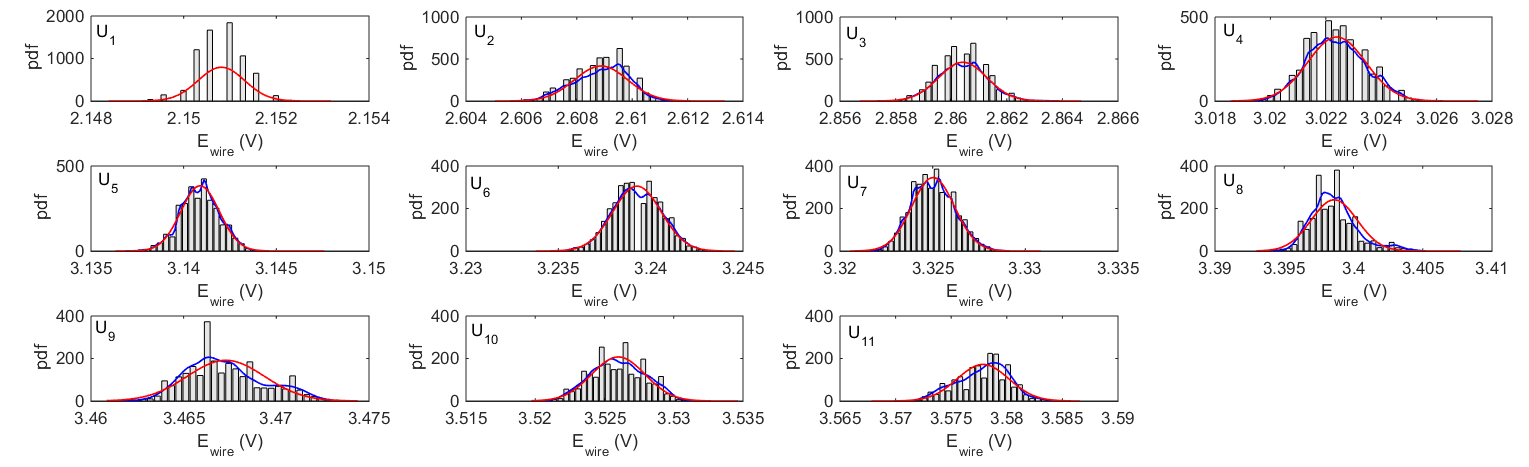}
\caption{Histograms and \revs{PDF (blue lines) of $10^4$ hot-wire voltage samples simultaneously measured with the Pitot-tube voltages of \fig ~\ref{fig:voltagePitotPdf} along with the closest Gaussian distributions fitted to the samples (red lines).} }\label{fig:voltageWirePdf}
\end{figure}

We start by explaining the propagation of uncertainties in the model shown in \fig ~\ref{fig:hwaUQ}, from the calibration voltages of the hot-wire anemometer and the Pitot tube.
In Figs. \ref{fig:voltagePitotPdf} and \ref{fig:voltageWirePdf}, the histograms of the voltage samples along with their associated PDFs are shown.
In addition to these, the closest Gaussian (normal) distributions that can be fitted to the data with the mean and variance equal to the ensemble ones are also plotted. 
Although for some flow conditions there is a relatively good agreement between the Gaussian and the sample distributions, \revf{some runs exhibit small gaps in certain bins. This is attributed to the relatively high wind-tunnel turbulence intensity, which increases the required number of samples to obtain smooth Gaussian distributions. In the most extreme cases, especially at higher flow velocities, the sample } PDF appears to be a combination of two or more Gaussian distributions. 
This implies that it is necessary to take even larger number of voltage samples when higher flow velocities are considered. 
However, \revf{for the purpose of illustrating the methodology and avoiding unnecessary complications, } this difference \revf{that is believed to have negligible impact on the overall process} is neglected in the upcoming analysis, relying on the assumption that if the number of samples increases then the distribution \revs{tends to} Gaussian, as discussed, for instance, in Ref. \cite{discetti}.

\begin{figure}[htb!]
\centering
\includegraphics[scale=0.53]{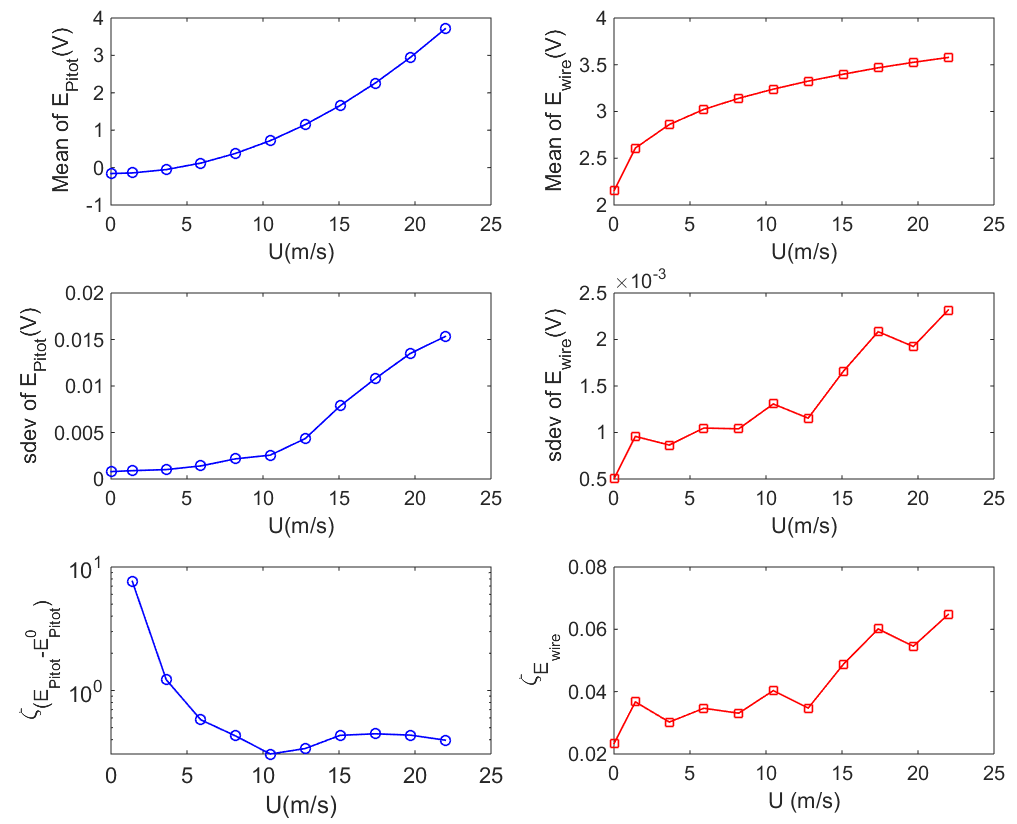}
\caption{\revs{Variation with freestream velocity of the ensemble mean, standard deviation, and the percentage of the uncertainty of the observed samples of (left) Pitot-tube and (right) hot-wire voltages, associated with the distributions shown in Figs. \ref{fig:voltagePitotPdf} and \ref{fig:voltageWirePdf}. To avoid negative values, the percentage of uncertainties of the Pitot-tube voltage samples shifted by no-flow data, \ie $(E_{\rm{Pitot}} - E^0_{\rm{Pitot}})$, is plotted.}}\label{fig:meanSdevVoltageData}
\end{figure}

Therefore, in order to explain the whole methodology, it is assumed that at each flow condition, the calibration voltages of the Pitot-tube and hot-wire have Gaussian distributions with the mean and variance equal to their corresponding ensemble ones. 
The mean, variance and covariance of the observed voltages are not \revs{constant, but vary} with \revs{the freestream} velocity\revs{, as shown in \fig ~\ref{fig:meanSdevVoltageData}}. 
In particular, the \revs{uncertainty in the Pitot-tube voltage samples is rapidly reduced with the flow velocity, up to $U\approx 10$\,m/s,} as opposed to $E_{\rm{wire}}$, for which the uncertainties increase \revs{monotonically} with the flow velocity.

In the next step, the calibration voltages of the Pitot tube are employed to calculate the corresponding PDFs of the dynamic pressure. This is then followed by plugging the outcoming distributions of $\Delta p$ along with the air temperature and the ambient pressure as uncertain inputs into equation (\ref{Ucal}) for evaluating distributions of the calibration velocity, which can then be written as $U_{\rm{Pitot}}$ in order to avoid ambiguities in the forthcoming analyses.

\begin{figure}[!t]
\centering
\includegraphics[scale=0.50]{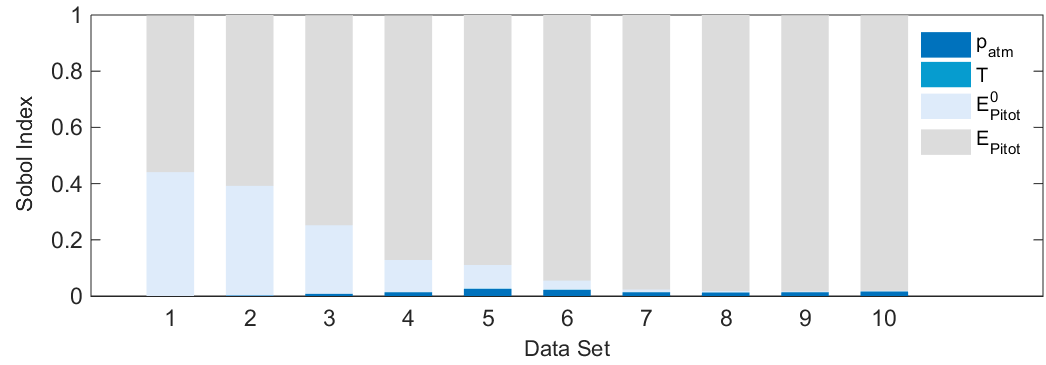}
\caption{Sobol indices of $U_{\rm{Pitot}}$ for all the 11 datasets used in the calibration process. }\label{fig:SobolCalibProcess}
\end{figure}

As discussed in \sect ~\ref{sec:fwd}, \revs{any of the perturbation, sample- and spectral-based methods} could be employed to assess the propagation of the uncertainties in $U_{\rm{Pitot}}$. \revs{However, in case of employing the sample- or spectral-based methods}, Sobol sensitivity indices of $U_{\rm{Pitot}}$ with respect to flow temperature, ambient pressure and Pitot-tube voltages can be calculated.
As shown in \fig ~\ref{fig:SobolCalibProcess}, $U_{\rm{Pitot}}$ is mainly sensitive to the Pitot-tube voltage readings\rev{, perhaps as expected}.
Moreover, as the flow velocity increases, the contribution of the Pitot-tube voltage at no-flow condition $E^0_{\rm Pitot}$ in the uncertainty of $U_{\rm{Pitot}}$ is gradually reduced, until it eventually vanishes.

Once the values of $U_{\rm{Pitot}}$ associated with the Pitot-tube calibration voltages are calculated through the above-described forward problem, it is required to use them along with the corresponding hot-wire calibration voltages to formulate a voltage--velocity calibration curve $E-U$. 
As explained in \sect ~\ref{sec:HWAmeasure}, there are conventional functional forms available for this curve and what is needed to be addressed is the estimation of the function parameters through solving a UQ inverse problem, \revs{with the techniques of} \sect ~\ref{sec:uqInv}. 
In particular, King's law (\ref{kings_law}) is considered here in the statistical models (\ref{eq:statisModelInverse}) and (\ref{eq:bayes_EiV}) with $E_{\rm{wire}}$ and $U_{\rm{Pitot}}$ \revs{corresponding to} $y$ and $x$, respectively. 
The aim is \revs{thus} to estimate the parameters $\theta=\{A,B,n\}$ given the observed data for calibration wire voltages \revs{and} Pitot-tube velocities. 

Due to the presence of uncertainties in both voltage and velocity data \revs{and also variation of the magnitudes of the uncertainties with the freestream velocity, see \fig ~\ref{fig:meanSdevVoltageData}}, the application of the ordinary LS method or the statistical model (\ref{eq:statisModelInverse}) seems to be questionable, at least at the first look. 
In fact, the LS method is a point estimator working based on minimizing the summation of the squares of the difference between the mean values of the observations and the mean value of the model response, while not taking into account the existence of the uncertainties involved in the observed data. 
In this framework\revs{, by employing the Bayesian model (\ref{eq:bayes}) with likelihood function (\ref{eq:GaussLikeli}) which is equivalent to the non-linear ordinary LS method}, the mean value of the model parameters along with an error \revs{variance} representing the model misfit is obtained. 

Contrary to these two \revf{methods}, if the sampled errors (in the form of a partial or full covariance matrix) in the wire voltage and Pitot-tube velocity at different \revs{freestream velocities} are used in conjunction with the error-in-variable Bayesian inference (\ref{eq:bayes_EiV}) with likelihood function (\ref{eq:likeliMultiGen}), then it is expected that the observed uncertainties in both velocity and voltages impact the King's law parameters in a more realistic \revs{way}.

\begin{figure}[!t]
\centering
\includegraphics[scale=0.42]{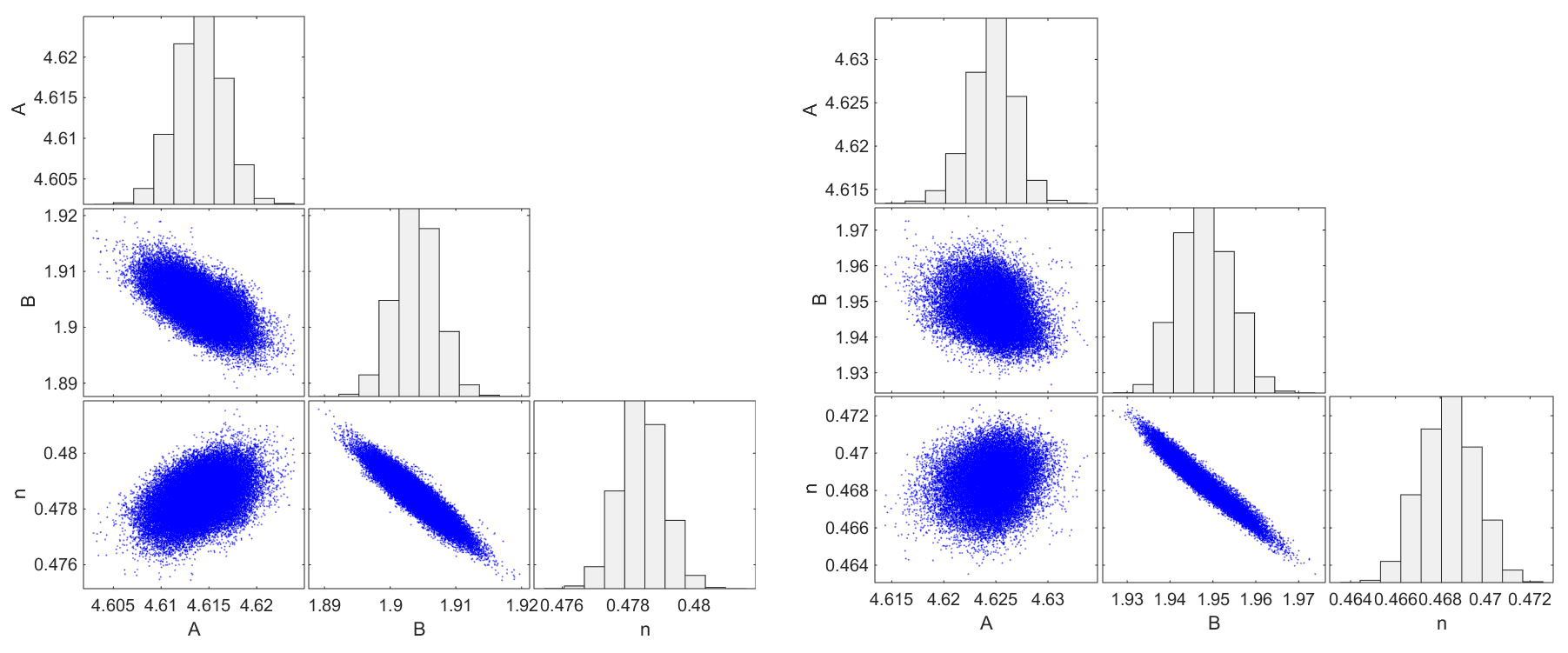}
\caption{Scattered and histogram plots of the King's law parameters estimated by Bayesian inference, (left) with and (right) without considering EiV model.}\label{fig:scatterKingsParams}
\end{figure}

For the specific datasets studied here, the mean and standard deviations of the King's law parameters calculated by different approaches along with the corresponding mutual covariances are listed in Tables \ref{tab:KingsParams} and \ref{tab:KingsParamsCovar}, respectively.
Although the mean values of the parameters \revs{calculated by} different methods are similar, implying that the produced curves will have values close to the mean values of the data, the uncertainty of the parameters determined by the non-linear LS method (and its Bayesian equivalence based on likelihood (\ref{eq:GaussLikeli}) with constant variance) is larger than those given by the Bayesian inference with \revs{varying} $\sigma^2_{y_j},\, j=1,2,\cdots,11$ in (\ref{eq:GaussLikeli}), denoted by Bayesian without EiV as well as the Bayesian with EiV model with likelihood function (\ref{eq:likeliMultiGen}). 
In the latter, the estimates of the elements of the covariance matrix (\ref{eq:covMatrixW}) are taken from the \revs{observed samples} in each of the 11 datasets.
The validity of these estimations relies on the fact that for each dataset, $10^4$ samples were taken during the experiment within short time intervals. 
\revs{The scatter plots and associated histograms of the King's law parameters estimated by Bayesian inference with and without considering EiV model are illustrated in \fig ~\ref{fig:scatterKingsParams}.}

A cautious conclusion \revs{which can be made by comparing different approaches for parameter estimation} is that, as more detailed information of the errors in the observed data is used in the inverse problem, the \revs{estimated} parameters become less uncertain. 
In other words, the uncertainty given by the LS method is fictitious due to the linearisation in constructing the covariance matrix for the estimated parameters and also assuming \revs{the iid constant-variance error $\varepsilon_i\sim \mathcal{N}(0,\sigma)$ for $i=1,2,\cdots,11$ in (\ref{eq:statisModelInverse})} to be able to reflect the correct structure of the uncertainties, \revs{while the variance and covariances of the observed data vary with freestream velocity.}

Another important observation is that according to the values given in Table \ref{tab:KingsParamsCovar}, the correlation coefficients between the parameters are not small, and as shown in \sect ~\ref{sec:ofi_hwaFWD}, they have a significant influence on the velocities predicted by the velocity--voltage calibration curve. 
Moreover, \revs{in the EiV model where covariances between the observed data} are taken into account in constructing (\ref{eq:covMatrixW}), the \revs{resulting} correlation in the estimated parameters becomes larger compared to the case without EiV modelling. 
\revs{This can be clearly observed from the scattered plots of the estimated parameters in \fig ~\ref{fig:scatterKingsParams}}.

To summarize the results thus far, the uncertainties in the observed calibration $E_{\rm{wire}}$ and $U_{\rm{Pitot}}$ data caused by various factors move forward to the model parameters of the $E-U$ calibration curve. 
Eventually, in the velocity--measurement phase of the experiment when the mentioned curve is used to predict the velocity associated to a hot-wire voltage reading (see Section \ref{sec:ofi_hwaFWD}), the uncertainties included in the model parameters become influential.

\begin{table}[!t]
\centering
\caption{Estimated values of mean and standard deviation of the King's law (\ref{kings_law}) parameters obtained using different methods.}\label{tab:KingsParams}
\begin{small}
\begin{tabular}{*{4}c}
\toprule\toprule
Approach & \multicolumn{3}{c}{Estimated Parameter Values} \\
{}& $A$ & $B$ & $n$\\
Ordinary Least Squares &4.6096$\pm$0.0370 & 1.9522$\pm$0.0397 & 0.4670$\pm$0.0063 \\
Bayesian with \revf{eq.} (\ref{eq:GaussLikeli}) & 4.6099$\pm$0.0380 & 1.9519$\pm$0.0411 & 0.4671$\pm$0.0066 \\
Bayesian without EiV & 4.6246$\pm$2.1918$\times 10^{-3}$ & 1.9481$\pm$6.0985$\times 10^{-3}$ & 0.4684$\pm$1.2062$\times 10^{-3}$ \\
Bayesian with EiV & 4.6140$\pm$2.5328$\times 10^{-3}$ & 1.9039$\pm$3.3081$\times 10^{-3}$ & 0.4784$\pm$6.7482$\times 10^{-4}$ \\
\hline
\end{tabular}
\end{small}
\end{table}

\begin{table}[!t]
\centering
\caption{Estimated values for the covariances and the associated Pearson's correlation coefficients of the King's law (\ref{kings_law}) parameters implementing different approaches. For two random variables $x$ and $y$, the Pearson's correlation coefficient is defined by $\rho_{xy}=\cov(x,y)/\sqrt{\var(x)\var(y)}$.}\label{tab:KingsParamsCovar}
\begin{small}
\begin{tabular}{*{7}c}
\toprule\toprule
Approach & \multicolumn{3}{c}{Covariance} & \multicolumn{3}{c}{Correlation Coefficient}\\
{}& $\cov(A,B)$ & $\cov(A,n)$ & $\cov(B,n)$ & $\rho_{AB}$ & $\rho_{An}$ & $\rho_{Bn}$\\
Ordinary Least Squares & -1.0934$\times 10^{-3}$ & 1.3252$\times 10^{-4}$ & -2.4078$\times 10^{-4}$ & -0.7441 & 0.5668 & -0.9584 \\
Bayesian with eq. (\ref{eq:GaussLikeli}) & -1.1492$\times 10^{-3}$ & 1.3957$\times 10^{-4}$ & -5.5877$\times 10^{-4}$ & -0.7359 & 0.5584 &  -0.9586 \\
Bayesian without EiV  & -4.0281$\times 10^{-6}$ & 5.0802$\times 10^{-7}$ & -7.1379$\times 10^{-6}$ & -0.3014 & 0.1922 & -0.9704\\
Bayesian with EiV & -5.3240$\times 10^{-6}$ & 7.5121$\times 10^{-7}$ & -2.0594$\times 10^{-6}$ & -0.6345 & 0.4395 &-0.9225 \\
\hline
\end{tabular}
\end{small}
\end{table}

\subsection{Accounting for the uncertainties in OFI measurements}\label{sec:ofiEx_factors}
As discussed in \sect ~\ref{sec:OFImeasure} and schematically illustrated in \fig ~\ref{fig:ofiUQ}, different factors including air temperature and ambient pressure, oil properties and optical parameters may affect the values of the wall-shear stress $\tau_w$, as shown by equations (\ref{Dh}) and (\ref{eq:tauOFI}). 
The aim of this section is to investigate the uncertainties associated with these factors.

In order to remain consistent in the analysis, we assume that $T$ and $p_{\rm atm}$ have \revs{the same} uncertainties as those considered in the previous section.
Among the optical parameters, $\lambda$ and $n_{\rm air}$ are taken to be fixed known parameters, while the camera angle $\alpha$ is prone to experience a $0.5^{\circ}$ uncertainty \revs{(assumed to be Gaussian)} for a nominal value of $15^{\circ}$.
The refraction index of oil, $n_{\rm oil}$ slightly varies with temperature.
According to Vinuesa et al. \cite{channel_ofi}, in the temperature range from $10^{\circ}$ to $50^{\circ}$\,C, $n_{\rm oil}$ varies about $0.15\%$ for a $200$ cSt oil, which is the one used in the experiment discussed in the current analysis. This small variation can be neglected, as for instance done in Ref. \cite{osterthes}. 
\revs{The uncertain air temperature and ambient pressure are assumed to be Gaussian and independent from each other and other parameters, with the mean and standard deviations given in Table \ref{tab:fwdOFI_ex1in}. }

Next, the focus will be on the study of the remaining factors which may also play significant roles. As described in \sect ~\ref{sec:OFImeasure}, the rate of change of the fringe wavelength, ${\rm d} \lambda_f / {\rm d} t$, can be calculated by means of the global wavelength estimation method, given observations of $\lambda_f$ as a time series.
For this purpose, the slope for $\lambda_f=c+ t \,{\rm d} \lambda / {\rm d} t $ can be determined by applying a Bayesian inverse technique.
In this case, we assume an additive error model with the Gaussian likelihood function (\ref{eq:GaussLikeli}) to estimate the parameters $c$, ${\rm d} \lambda_f / {\rm d} t$, and the variance of the error between the observed data and the linear fit. 
A crucial issue here is choosing the range of data in the $\lambda_f - t$ series to perform the fit. 
Examining different data batches reveals a considerable impact on the standard deviation of ${\rm d} \lambda_f / {\rm d} t$ with less remarkable influence on the associated mean value. 
In particular, when the \revCom{later} images in the time series are selected, an estimate for ${\rm d} \lambda_f / {\rm d} t$ with larger standard deviation (up to $1.2\%$ of the mean value) is obtained. 
In contrast, by selecting observation data at lower times, where the fringes are less scattered, an average mean and standard deviation of $4.34210 \times 10^{-6}$ \revs{\,m/s} and $7.46882 \times 10^{-9}$ \revs{\,m/s}, respectively, \revs{are obtained}. These will be used in the next section. 
It must be noted that the posterior density of ${\rm d} \lambda_f / {\rm d} t$ is Gaussian because of the specific choice of the likelihood function.

\begin{figure}[!t]
\centering
\includegraphics[scale=0.60]{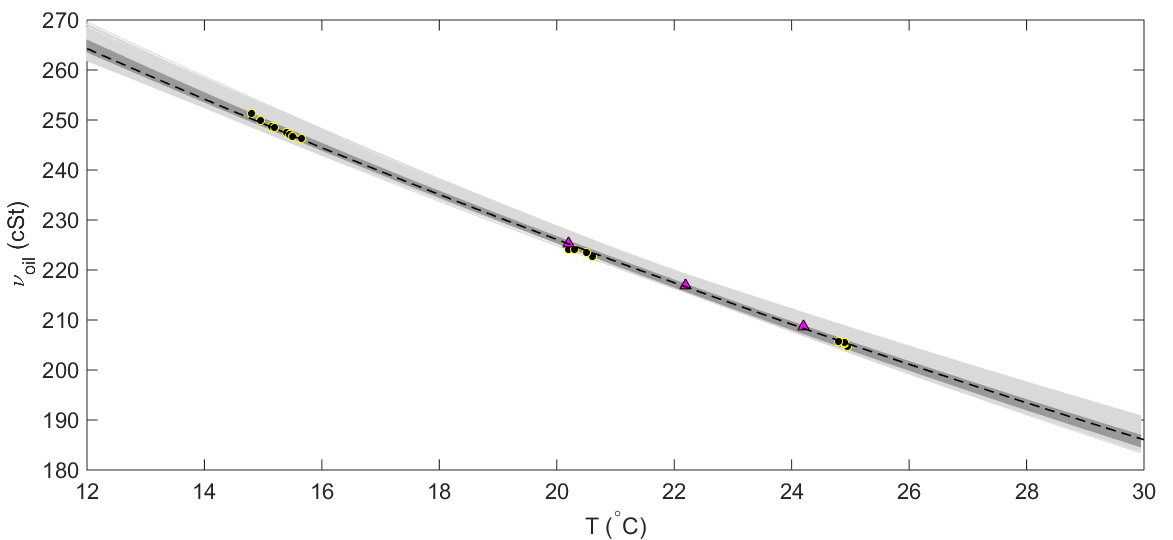}
\caption{Variation of $\nu_{\rm{oil}}$\,(cSt) with $T$\,($^\circ$C) given by eq. (\ref{eq:oilVistT}), along with the dashed line representing the curve proposed by \"Osterlund \cite{osterthes}. 
\revf{The values in the light and dark shaded areas represent predictions of eq. (\ref{eq:oilVistT}), when, respectively, $1\%$ and $5\%$ error is involved in the observed data of temperature.} The experimental datasets were reported by Vinuesa et al. \cite{channel_ofi}. The two datasets reported in that study, namely IIT-2009 and EPFL-2008, are represented by $\tiny{\bigtriangleup}$ and $\bullet$, respectively.
}\label{fig:nuOilFit}
\end{figure}

An important factor contributing to the uncertainty of the wall-shear stress measured by means of OFI is the oil viscosity. 
As stated above, the variation of oil density with temperature is neglected, which makes the uncertainties in the kinematic and molecular oil viscosities linearly proportional. 
According to \"Osterlund \cite{osterthes}, the functional form:
\begin{equation}\label{eq:oilVistT}
\nu_{\rm{oil}}= a_\nu \exp \left [b_\nu \left (25^{\circ} {\rm C} -T \right ) \right ] 
\end{equation}
is appropriate to express the variation of the oil viscosity with temperature. 
For known parameters $a_\nu$ and $b_\nu$, this relation can be used to calculate the oil viscosity in an OFI experiment within the forward problem. 
Therefore, it is required to perform an inverse problem in advance to determine the values of the model parameters $\theta=\{a_\nu,b_\nu\}$ given a set of separately observed data \revs{$(\nu_{\rm{oil}},T)$} by means of a viscometer, a device which has a nominal accuracy of $0.1\%$ that might \revs{change to} up to $0.3\%$ in practice \cite{vinuesa_orlu,channel_ofi}.
Since in addition to the oil viscosity, the independent variable in (\ref{eq:oilVistT}), {\it i.e.} $T$, is subjected to measurement uncertainties, the error-in-variable model (\ref{eq:bayes_EiV}) can be implemented along with the likelihood function (\ref{eq:likeliMultiUncor}). 
However, due to the small error in temperature of $0.1\%$, the inclusion of the uncertainty in $T$ in addition to that of $\nu_{\rm{oil}}$ does not affect the mean and variance of the estimated parameters $a_\nu$ and $b_\nu$, a fact that justifies using a non-linear LS method. 
To assess the procedure, a curve in the form of expression (\ref{eq:oilVistT}) is fitted to the experimental data shown in Fig. \ref{fig:nuOilFit}. 
The mean and standard deviations of the coefficients are determined to be $a_\nu=204.7676\pm 0.1468$\revs{\,cSt} and $b_\nu=0.01972 \pm 1.03850 \times 10^{-4}$\revs{\,$^\circ$C$^{-1}$} for $0.1\%$ Gaussian error in $T$. 
If the error in $T$ increases to $1\%$, then $\nu_{\rm oil}$ would be within the range specified by the dark shaded area in \fig ~\ref{fig:nuOilFit} which includes $\nu_{\rm{oil}}=205.1\exp[0.0195(25^{\circ} {\rm C} -T)]$, the curve reported by \"Osterlund \cite{osterthes}.

In order to see the role of the temperature errors in calibrating (\ref{eq:oilVistT}), let us assume an extreme case in which the error in the observed temperatures is $5\%$. 
By employing the Bayesian inference with EiV model (\ref{eq:bayes_EiV}) with likelihood (\ref{eq:likeliMultiUncor}), the values of parameters are estimated to be $a_\nu=206.0446 \pm 0.7745$\revs{\,cSt} and $b_\nu=0.01930\pm 4.4068\times 10 ^{-4}$\revs{\,$^\circ$C$^{-1}$}. 
This increase in the parameter uncertainty would consequently result in an uncertain prediction of $\nu_{\rm{oil}}$ shown by the light shaded area in \fig ~\ref{fig:nuOilFit}.
It is important to note that in any case, the minimum uncertainties \revs{in the predictions exist in the interval within which the observed data are available}. This requires exercising extra care when a calibrated oil viscosity-temperature curve is being used to predict $\nu_{\rm{oil}}$ outside of the range of the observed data for temperature.

\section{Propagation of the uncertainties in the QoIs of HWA and OFI}\label{sec:ofi_hwaFWD}

The previous sections provided the details on how to compute the uncertainties originated from the different sources as outlined in Figs.~\ref{fig:hwaUQ} and~\ref{fig:ofiUQ}. 
The propagation of these uncertainties into the QoIs, which are the flow velocities at specific distances from the wall measured by HWA, and the wall-shear stress obtained by means of OFI, can be evaluated using any of the approaches described in \sect ~\ref{sec:fwd} for the forward problem. In particular, we decided to expand each QoI in terms of the involved parameters via equation (\ref{eq:pceExpans}). 
Then, the mean and variance of the QoIs can be constructed through equations (\ref{eq:pceMean}) and (\ref{eq:pceVar}). 
For the parameters with Gaussian distribution, Hermite polynomials are used in the PCEs, which according to Ref. \cite{xiu_gPCE} are the optimal choice. 
To find the coefficients of the expansions, the Smolyak sparse grid integration method \cite{smolyak} with sparse grid levels 3 \revs{and 5 as well as the tensor-product quadrature method, see \eg \cite{smith}, with quadrature orders 3 and 5 in each parameter space dimension as already available} in the Dakota library \cite{dakotaMan}, are employed. 
In case of using PCEs, the main and total global sensitivity indices with interaction level 2 can be evaluated analytically. 
Alternatively, the distributions of the QoIs can also be directly constructed by a Monte Carlo sampling method.
In such a case, the calculation of the Sobol indices is done by direct evaluation of equations (\ref{eq:sobolMain}) and (\ref{eq:sobolTotal}), which is however computationally expensive. 
In producing the following numerical results, both PCE and LHS Monte Carlo approaches along with the deterministic perturbation method (\ref{eq:varPerturb}) were \revs{all} employed. 
The interesting conclusion is that the three approaches predict almost the same mean value of the QoIs. 
The difference between the variances is also found to be \revf{comparably} small. 
This may legitimate using the classical approach, equation (\ref{eq:varPerturb}), for the analysis of the current problem, motivating the negligible non-linearity existing in the expressions relating the inputs to the responses. 
However, in order to find the Sobol indices, it is still required to use a MC-based approach or the PCE method, where the latter was the main choice due to its computational efficiency and accuracy. For all the numerical results produced here, the following parameters are considered \revs{to be \revf{known and} fixed}: air gas constant $R=287.0$\, J/kg$\cdot$K, air and oil refraction indices $n_{\rm air}=1.0$ and $n_{\rm oil}=1.4032$, oil density $\rho_{\rm oil}=967.0$ kg/m$^{3}$, and wavelength of the Sodium lamp $\lambda=589.3$\,nm. 
We assume the uncertainty in ambient pressure and air temperature to be, respectively, $0.1\%$ and $0.2\%$ in the $95\%$-confidence sense. 
The mean and standard deviations of these two quantities along with the uncertain parameters in the OFI as quantified in \sect ~\ref{sec:ofiEx_factors} are listed in Table \ref{tab:fwdOFI_ex1in}.

\begin{table}[!t]
\centering
\caption{\revs{Input} values of $p_{\rm atm}$, $T$, and parameters of OFI to perform the uncertainty propagation.}\label{tab:fwdOFI_ex1in}
\begin{small}
\begin{tabular}{*{7}c}
\toprule\toprule
{} & \multicolumn{2}{c}{General}& \multicolumn{4}{c}{OFI} \\
Input Quantity & $p_{\rm atm}$ & $T$ & $a_\nu $ & $b_\nu$  & ${\rm d} \lambda_f / {\rm d} t$ & $\alpha $ \\
Units & $\rm(Pa)$ & $\rm (^\circ C)$ & $\rm (cSt)$ &$\rm (1/^\circ C)$ & $\rm (m/s)$ & $\rm (deg)$ \\
Mean &100700.0 & 22.1 & 204.7676 & 0.01972  & 4.34210$\times 10^{-6}$ &  15\\
sdev &50.0 & 0.0255 & 0.1468  & 1.03850$\times 10^{-4}$ & 7.46882$\times 10^{-9}$ & 0.5 \\
\hline
\end{tabular}
\end{small}
\end{table}

\begin{figure}[!t]
\centering
\includegraphics[scale=0.5]{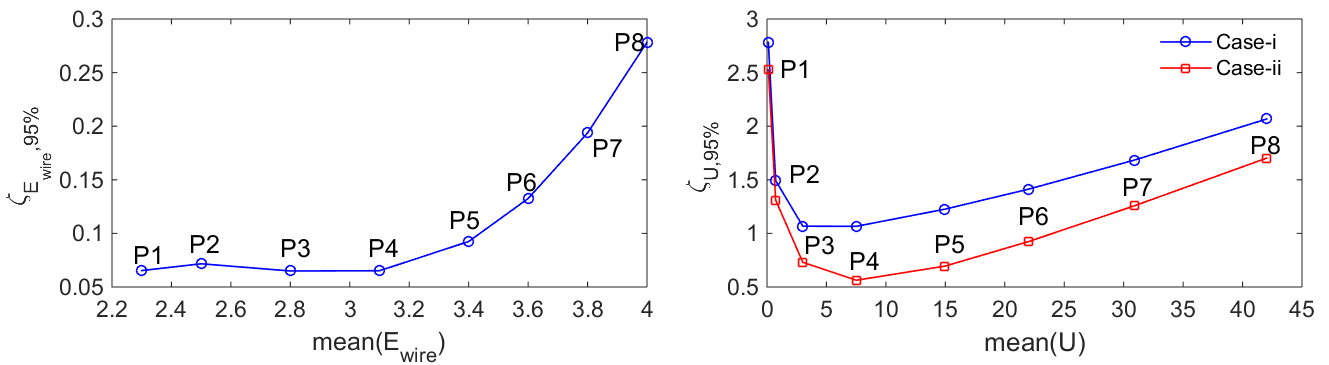}
\caption{\revs{(Left) Considered mean and error percentage (in 95\%-confidence sense) of wire voltages as inputs to the forward problem performed for HWA. (Right) Resulting mean and error percentages (in 95\%-confidence sense) of velocity.} Note that in Case-i the correlations between the King's law parameters are ignored, whereas in Case-ii the mutual correlations between them are taken into account.}
\label{fig:HWA_FWD_8Cases}
\end{figure}

To obtain the flow velocity corresponding to a certain hot-wire voltage, the velocity--voltage calibration curve (\ref{kings_law}) can be employed. Note that for this curve the model parameters estimated in the calibration phase are presented in Tables \ref{tab:KingsParams} and \ref{tab:KingsParamsCovar}.
\revs{In particular, we use the parameter values estimated by the Bayesian inference with EiV model, since as thoroughly discussed in \sect ~\ref{sec:hwaEx_factors}, the associated uncertainties of these parameters are more realistic.} 
In order to investigate how the correlation between these model parameters may influence the results, two different cases are considered. 
In the first one, denoted as Case-i, the correlations among parameters are ignored. This implies that the parameters are mutually independent. In the second case, denoted as Case-ii, the mutual correlations estimated via Bayesian inference with EiV model (\revs{see Table \ref{tab:KingsParamsCovar}}) are taken into account. At a specific distance from the wall, voltage readings have certain distributions that can be \revf{assumed} to be Gaussian, provided that enough samples are gathered. 
Here, we consider \revs{8 different} mean values of hot-wire voltages $\bar{E}_{\rm wire}$ larger than $2.15$\,V, with the associated standard deviation of the signals to be given by $(0.23480 \bar{E}^3_{\rm wire}-1.92268\bar{E}^2_{\rm wire}+5.24295\bar{E}_{\rm wire}-4.66791)\times 10^{-2}$\revs{\,V}.
This relation is derived based on the observed data in this particular example. 
The mean and \revs{95\%-confidence error} of the \revs{8} specific wire voltages considered here are \revs{shown in \fig ~\ref{fig:HWA_FWD_8Cases} (left).} The velocities corresponding to these voltage readings as the outcome of the forward problem conducted by including the uncertainties due to the $E-U$ calibration curve and voltage readings for Case-i and Case-ii are \revs{plotted in \fig ~\ref{fig:HWA_FWD_8Cases} (right).} 

Let us recall that the parameters are $q=\{A,B,n\}$, and whereas in Case-i the correlations between them are not taken into account, in Case-ii we also consider them. \fig ~\ref{fig:HWA_FWD_8Cases} (right) shows that in both Case-i and Case-ii, the relative error of the velocity exhibits a particular trend with the mean velocity: the largest error is observed at the lower velocities, then the error decreases rapidly until reaching a minimum in Case P4. Furthermore, the higher-velocity cases exhibit a slight increase in error with flow velocity, a fact that could be related to an insufficient number of samples in the higher-velocity datasets. Note that the HWA and OFI measurements analyzed in the present work were extracted from different experiments. However, if the OFI measurements summarized in Table \ref{tab:fwdOFI_ex1in} were obtained in the same experiment as the HWA measurements, a similar pattern would be observed for $U^{+}$. This may be of a particular interest when the observations for the velocity and inner-scaled wall-normal location $y^{+}=y u_{\tau} / \nu$ are used to estimate the parameters of the overlap region in turbulent boundary layers. In such a case, to estimate the uncertainties in the parameters as accurately as possible, it would be necessary to construct the likelihood function in conjunction with the Bayes' inverse theorem in a way that it reflects a realistic picture of the contributing uncertainties in the $U^{+}$, $y^{+}$ data.

It is also noteworthy that, \revs{as shown in \fig ~\ref{fig:HWA_FWD_8Cases} (right),} the estimated errors in the measured velocities in Case-ii are lower than those of Case-i, although the mean velocities are the same in both cases. 
This implies that when the King's law parameters are considered to be correlated, the resulting error in the estimated velocity is reduced, an outcome that may seem unexpected. But, relying on the fact that the \revs{covariance matrix} of the response can be approximated by equation (\ref{eq:varPerturb}), it is possible that the summation of the sensitivity coefficients multiplied by the mutual correlations among parameters leads to a reduction in the effect of the variance of the individual parameters (when considered uncorrelated). This is an important result, since in most studies of the experimental uncertainty in wall-bounded turbulence, the correlations among parameters are not taken into account when determining the overall uncertainty of a measured quantity. The fact that Case-ii shows lower error levels than those of Case-i indicates that the mutual interactions among parameters may lead to cancellation of errors, and thus to a lower overall uncertainty.

\begin{figure}[t!]
\centering
\includegraphics[scale=0.50]{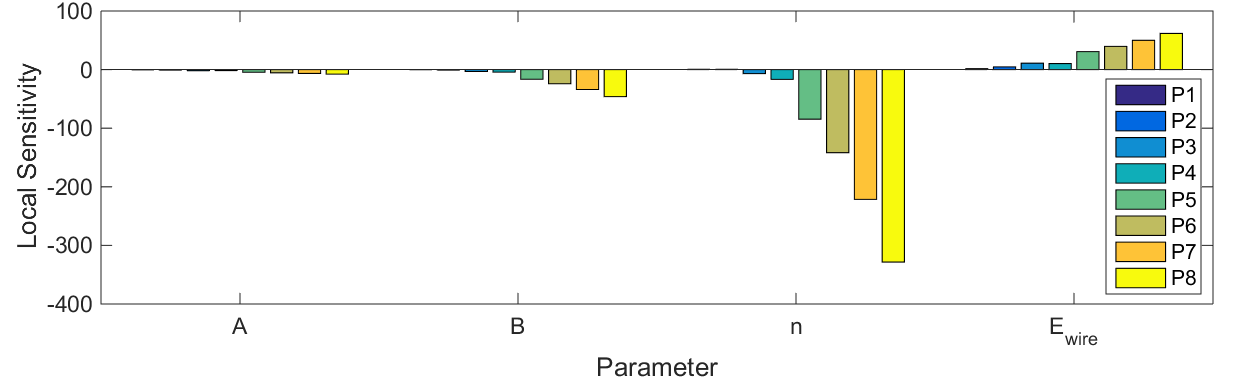}
\caption{Local sensitivity indices for the measured velocity $U$ \revs{with respect to the King's law parameters and hot-wire voltage.} Note that in this local analysis the parameters are perturbed around their nominal values.}\label{fig:HWA_LSAindices}
\end{figure}

For the same set of simulations, local and global sensitivity analyses can be performed. 
It is recalled that the local sensitivity indices illustrate the sensitivity of the response when the parameters are perturbed around their nominal values, which can be taken to be their mean. As shown in \fig ~\ref{fig:HWA_LSAindices}, the local sensitivity indices reveal that when the parameters are perturbed around their nominal values, the flow velocity is most sensitive to the change of parameter $n$ in the King's law (\ref{kings_law}) than the other parameters. 
This figure also shows that the local sensitivities with respect to all parameters increase monotonously with the flow velocity. 

\begin{figure}[t!]
\centering
\includegraphics[scale=0.50]{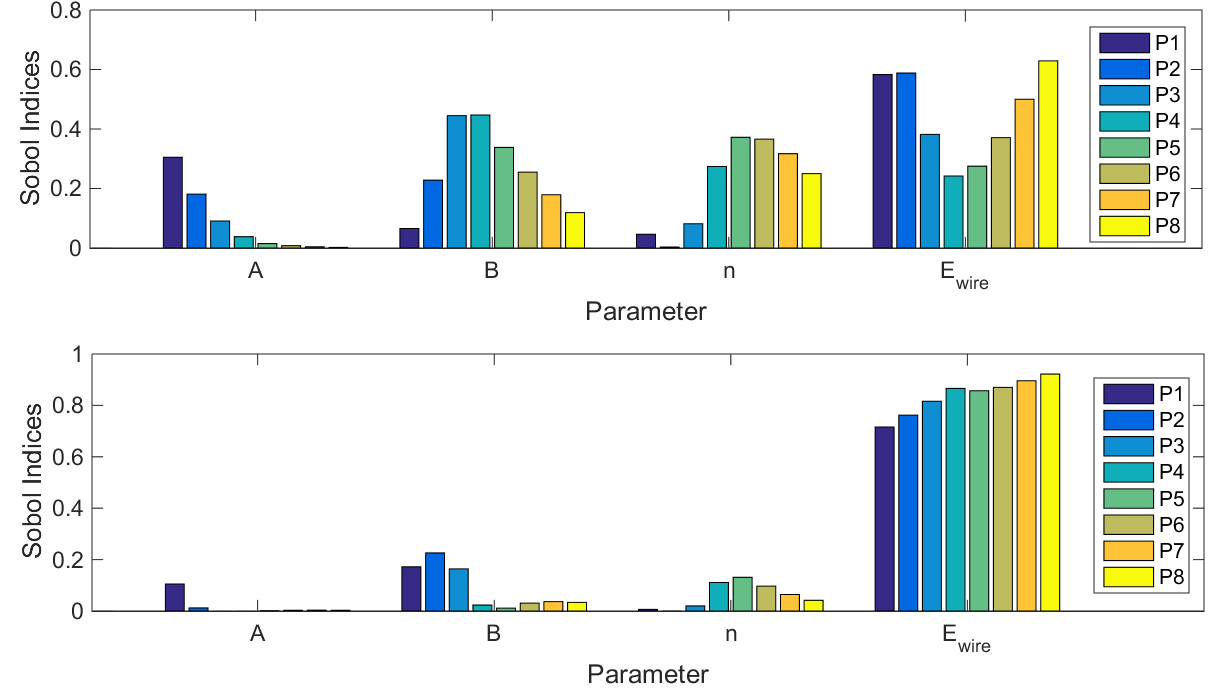}
\caption{Sobol sensitivity indices for the measured velocity $U$ (top) when all parameters are considered uncorrelated \revs{(Case-i)} and (bottom) when $A$, $B$, and $n$ are correlated \revs{(Case-ii)}. In the global sensitivity analysis, all the parameters are allowed to vary over their whole admissible space.}\label{fig:HWA_GSAindices}
\end{figure}

On the other hand, the global sensitivity analysis in the form of Sobol indices \cite{sobol} lead to the the results in \fig ~\ref{fig:HWA_GSAindices}. Whereas in the local analysis the parameters were perturbed around their nominal values, in the GSA the sensitivity of the response is studied while these parameters vary over their whole admissible space. Doing so, it is possible to infer how much each parameter contributes to the uncertainty of the resulting response. According to the results shown in \fig ~\ref{fig:HWA_GSAindices}, it is clear that the inclusion or exclusion of the correlation between the King's law parameters significantly affects the values of the Sobol indices of the flow velocity. In particular, when the correlations among parameters are considered (Case-ii), the most sensitive parameter is the wire voltage regardless of the flow velocity. The Sobol indices of the wire voltage are approximately 9 times larger than the ones of the parameter $n$ in King's law, a fact that highlights the extreme sensitivity of the measured velocity to the wire voltage. On the other hand, when the correlations among parameters are ignored (Case-i), the relative sensitivity to the wire voltage decreases with respect to the King's law parameters. At low velocities the parameter $A$ and the wire voltage are the quantities with the highest Sobol indices. At intermediate velocities (where according to \fig ~\ref{fig:HWA_FWD_8Cases}, the relative error was lowest), parameters $B$ and $n$ are the ones with highest indices, even larger than the wire voltage. At higher velocities the behavior is similar to the one observed in Case-ii, but the relative difference between the wire voltage and the King's law parameters is significantly reduced. Combining the information conveyed by the global sensitivity analysis with the errors in the response, here the flow velocity, it is possible to obtain hints to reduce the uncertainties. 
For instance, in order to reduce the uncertainty in the velocity of Case P1, it is necessary to reduce the uncertainty of the wire voltage reading (considering that the other influential factors are kept unchanged).

\revCom{
Note that for convenience, and due to the relatively limited number of available flow measurements, here we considered King's law for the hot-wire calibration. 
However, it is common to use higher-order polynomials for this calibration, in particular fourth-order polynomials are widely used for turbulence measurements. 
Given that sufficient flow measurements were available to accurately obtain the polynomial coefficients, the Sobol indices associated to the wire voltage in \fig~\ref{fig:HWA_GSAindices} would remain approximately unchanged, and the overall contribution of the calibration parameters would be similar to that of the combined contributions of $A$, $B$ and $n$.
}

\begin{figure}[!t]
\centering
\includegraphics[scale=0.55]{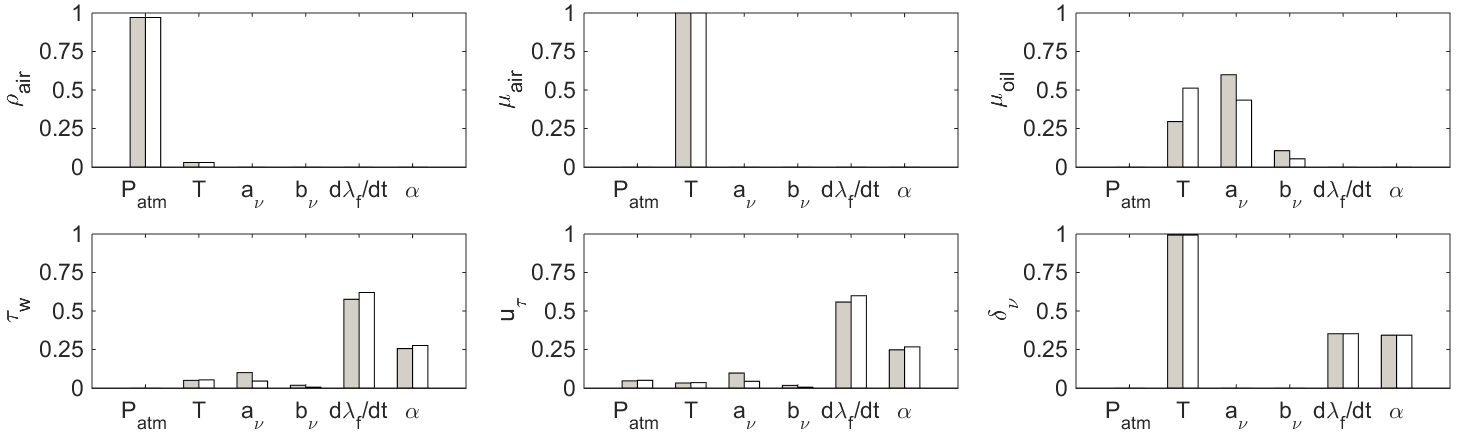}
\caption{Total Sobol sensitivity indices for different QoIs in GSA of OFI, with respect to various uncertain parameters given by Table \ref{tab:fwdOFI_ex1in}. \revs{Grey} indicates cases where all the parameters are considered uncorrelated and white the ones where $a_\nu$ and $b_\nu$ are correlated.}\label{fig:OFI_sobol}
\end{figure}

Similarly, \fig ~\ref{fig:OFI_sobol} shows the Sobol indices for the parameters in the GSA of the OFI experiment associated with \revs{the inputs in} Table \ref{tab:fwdOFI_ex1in}. Specifically, for the wall-shear stress $\tau_{w}$ and friction velocity $u_{\tau}$, \revs{the rate of change of the fringe wavelength}, and to less extent the camera angle and the coefficient $a_\nu$ in (\ref{eq:oilVistT}), have the most significant contributions. 
As opposed to the HWA measurements, inclusion of the correlations between the parameters in (\ref{eq:oilVistT}) does not significantly change the sensitivity indices. 
The most significant effect of \revs{including the correlations among parameters} is observed for the oil viscosity, where the sensitivity with respect to $a_\nu$ decreases with a corresponding increase of the temperature index. 
It is also important to note that, for the viscous length scale $\delta_\nu = \nu / u_{\tau}$, the summation of all the Sobol indices is larger than one. 
This is due to the fact that the temperature affects both $\nu$ and $u_\tau$, simultaneously. 
Note that from all the variables studied here, this is the only case in which the main and total Sobol indices are different. 

The mean and uncertainty percentages of the QoIs of OFI, corresponding to the inputs of Table \ref{tab:fwdOFI_ex1in}, are summarized in Table \ref{tab:UQHW_ex1out}. According to the results shown in this Table, the relative errors in the three most relevant quantities measured in OFI are: $0.44\%$ in the wall-shear stress $\tau_{w}$, $0.23\%$ in the friction velocity $u_{\tau}$ and $0.22\%$ in the viscous length $\delta_{\nu}$. It is interesting to note that these values are lower than those reported in other OFI studies in the literature. For instance, Vinuesa et al. \cite{channel_ofi} reported uncertainties of $0.85\%$ and  $0.58\%$ in $\tau_{w}$ and $u_{\tau}$, respectively; Segalini et al. \cite{ofi_segalini} reported an uncertainty of around $0.5\%$ in $u_{\tau}$; and Bailey et al. \cite{bailey} documented a significantly larger uncertainty in $\delta_{\nu}$ of $1.9\%$. An explanation for this discrepancy lies in the fact that in the present work we accounted for the interactions among parameters, which can effectively lead to a mutual cancellation of errors and therefore to a lower overall uncertainty. This was observed also in Fig.~\ref{fig:HWA_FWD_8Cases}, where the uncertainty in velocity measurements obtained by considering the interactions among parameters (Case-ii) was lower than the one where those interactions were neglected (Case-i).

\begin{table}[!t]
\centering
\caption{Mean and standard deviation as well as error percentage (in $95\%$-confidence sense) of QoIs and responses as a result of propagation of the uncertainties given in Table \ref{tab:fwdOFI_ex1in}.}\label{tab:UQHW_ex1out}
\begin{tabular}{*{4}c}
\toprule\toprule
QoI & Mean  & \revs{$\zeta_{95\%}$}\\
\hline
$\rho_{\rm air}$\,(kg/m$^3$) & 1.1884 &	 9.88$\times 10^{-2}$ \\
$\mu_{\rm air}$\,(kg/m$\cdot$s) & 1.1410$\times 10^{-6}$ &  5.12$\times 10^{-1}$\\
$\mu_{\rm oil}$\,(kg/m$\cdot$s) & 2.0966$\times 10^{-1}$ &	 1.82$\times 10^{-1}$\\
$\tau_{w}$\,(N/m$^2$)	 		  & 4.2610 &	 4.44$\times 10^{-1}$ \\
$u_\tau$\,(m/s)            & 1.8936	&  2.26$\times 10^{-1}$\\
$\delta_\nu$\,(m) 	  & 5.0997$\times 10^{-7}$ &  2.16$\times 10^{-1}$\\
\hline
\end{tabular}
\end{table}

\subsection{Estimation of uncertainty propagation in inner-scaled flow velocity}

\begin{figure}
\centering
\includegraphics[scale=0.50]{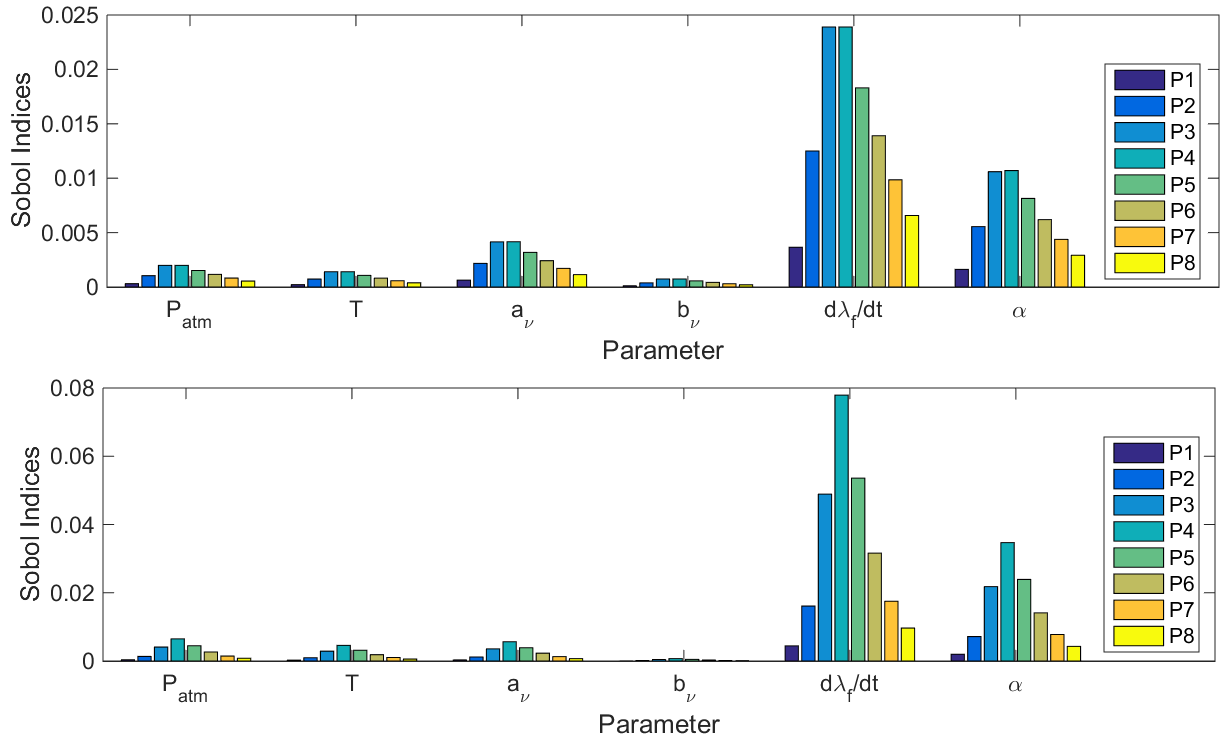}
\caption{Sobol sensitivity indices for $U^+$ from OFI measurements. (Top) All the parameters are considered uncorrelated \revs{ (Case-i)}, and (bottom) $A,\, B,\,  n$ as well as $a_\nu ,\, b_\nu$ are correlated \revs{ (Case-ii)}. Note that the Sobol sensitivity indices for $U^{+}$ from the HWA parameters correspond to the ones shown in Fig.~\ref{fig:HWA_GSAindices}.}\label{fig:allSobolUpls_ofi}
\end{figure}

As discussed above, the HWA measurements of flow velocity and OFI determination of \revf{the} wall-shear stress were carried out in different experiments, and therefore a comprehensive analysis of the uncertainties in $U^{+}$ cannot be carried out with these sets of data. However, in this section we aim at identifying the key parameters when determining such an uncertainty. To this end, we performed the GSA of the $U^{+}$ measurements that would be obtained from the HWA and OFI results discussed in this work. Note that the velocity measurements were carried out in a wind tunnel for teaching purposes, and therefore the flow quality is not the one that would be obtained in a typical wind tunnel with highly-controlled conditions used in wall-bounded turbulence research. For the GSA of $U^{+}$, we will consider the parameters analyzed for $U$ ($A$, $B$, $n$ and $E_{{\rm wire}}$), together with the ones investigated for $\tau_{w}$ ($p_{{\rm atm}}$, $T$, $a_{\nu}$, $b_{\nu}$, ${\rm d} \lambda_{f} / {\rm d}t$ and $\alpha$). In such a global sensitivity analysis, the Sobol indices for $U$ shown in Fig.~\ref{fig:HWA_GSAindices} would directly correspond to the ones for $U^{+}$ associated with those four parameters. Moreover, the Sobol indices for $U^{+}$ with respect to the OFI measurements are shown in Fig.~\ref{fig:allSobolUpls_ofi}. Thus, the complete estimated GSA for $U^{+}$ can be obtained from Figs.~\ref{fig:HWA_GSAindices} and \ref{fig:allSobolUpls_ofi}. By comparing the relative ranges of the Sobol indices from both figures it can clearly be observed that the estimated $U^{+}$ measurement is significantly more sensitive to the $U$-related quantities than to the ones associated with the wall-shear stress. This is of course expected, due to the different scope of both experiments. In particular, if the correlations among parameters are considered (Case-ii), the highest sensitivity is observed with respect to the wire velocity in all the flow velocities, with values of the Sobol indices on the order of 10 times larger than those of ${\rm d} \lambda_{f} / {\rm d}t$. The Sobol indices associated to $B$ and $n$, lower than those from $E_{{\rm wire}}$, are also around 10 times larger than the Sobol indices of the second most sensitive quantity from the OFI measurements: the camera angle. A similar behavior is noticed when neglecting the correlations among variables (Case-i), where it is interesting to note that in the case with lowest wire voltage Sobol index (case P4) the indices from fringe velocity and camera angle are largest (although still an order of magnitude lower). Therefore, although the OFI measurements were extremely accurate, the overall uncertainty of the $U^{+}$ measurements is relatively high due to the flow conditions established in the wind tunnel. This analysis, together with the choice of this particular HWA dataset, highlight the importance of the wind-tunnel flow conditions when performing accurate measurements of wall-bounded turbulence.

\section{Summary and concluding remarks} \label{summary_conclusions}

In the present work we evaluated the sources of uncertainty in two widely used measurement techniques in wall-bounded turbulence research, namely hot-wire anemometry (HWA) for flow velocity and oil-film interferometry (OFI) for wall-shear stress. To this end, both statistical and classical methods were employed for the forward and inverse problems to assess the contributions to the uncertainty of the various  components of the measurements. As may be expected, the non-linear least-squares method and the Bayesian inference \revs{with different likelihood functions} led to almost identical estimations of the mean values of the model parameters. 
\revs{However, the Bayesian inference with error-in-variable (EiV) model in which the likelihood function is constructed in accordance with the structure of the errors in the observed data, estimates the model parameters to be less uncertain.} Note that in the forward problem, \revf{only} small differences were found between applying Monte Carlo (MC), Polynomial Chaos Expansion (PCE) and linear perturbation methods. 
\revs{This \revf{supports} the validity of the classical perturbation method for studying uncertainty propagation in HWA-OFI measurements, despite its underlying simplicity.}

\revs{In addition,} the following conclusions were obtained after performing a propagation of uncertainty analysis for the HWA and OFI measurements: 
As observed in \fig~\ref{fig:HWA_FWD_8Cases}, the relative error in flow measurements is largest at low velocities, and reaches a minimum at intermediate velocities. The observed \revCom{rise} at higher velocities is indicative of the need for a larger number of samples with increasing flow velocity. Moreover, if the correlations among parameters are considered, the overall error is lower than if such correlations are neglected. Note that in most studies of wall-bounded turbulence these correlations are not considered, and the uncertainties from the various parameters are directly added when determining the overall uncertainty of the measured quantity.

A local sensitivity analysis of the flow velocity, in which the parameters are perturbed around their nominal values, reveals that the exponent in King's law $n$ is the parameter with largest influence on the overall uncertainty (see \fig ~\ref{fig:HWA_LSAindices}). However, if a global sensitivity analysis (in which the parameters are varied over their whole possible range) is carried out, then the measured wire voltage is the variable with the most prominent contribution as shown in \fig~\ref{fig:HWA_GSAindices}. Moreover, if the correlations among variables are also considered, the wire voltage becomes even more dominant in terms of contribution to the overall uncertainty of $U$.

Similarly, a global sensitivity analysis performed on the OFI measurements (see \fig~\ref{fig:OFI_sobol}) revealed that the factors contributing the most to the overall uncertainty in wall-shear stress measurements are the calculation of fringe velocity and the camera angle. In this case, accounting for the correlations among parameters does not significantly change the Sobol indices. Regarding the viscous length, the quantity with largest contribution to the overall uncertainty is the temperature measurement.
\revCom{
It is important to note that the most significant influence of the temperature on the OFI measurements is through its impact on the oil viscosity. This highlights the relevance of obtaining accurate viscosity calibration curves, as stated for instance by R{\"u}edi \et~\cite{ruedi}.
}

The present uncertainty propagation analysis shows that the relative errors in $\tau_w$, $u_\tau$ and $\delta_\nu$ are $0.44\%$, $0.23\%$ and $0.22\%$, respectively, as shown in Table \ref{tab:UQHW_ex1out}. 
These errors are lower than the ones reported in other studies in the literature, where the correlations among parameters are not considered. 
\revCom{To provide some examples, uncertainties of $0.85\%$ and $0.58\%$ were reported by Vinuesa \et~\cite{channel_ofi} for $\tau_{w}$ and $u_{\tau}$, respectively; Segalini \et~\cite{ofi_segalini} obtained an uncertainty of around $0.5\%$ in $u_{\tau}$ and a larger uncertainty in $\delta_{\nu}$, $1.9\%$, was reported by Bailey \et~\cite{bailey}.}
These results highlight the importance of accounting for such correlations when determining the overall uncertainty.

An approximate global sensitivity analysis was also performed for $U^{+}$, and as shown in Figs.~\ref{fig:HWA_GSAindices} and \ref{fig:allSobolUpls_ofi} the Sobol index corresponding to the wire voltage is around 10 times larger than the one associated with the calculation of the fringe velocity (which is the most sensitive parameter from OFI). This is due to the fact that, in this particular study, the $U$ measurements were obtained in a wind tunnel used for teaching purposes and therefore the flow quality is not suitable for wall-bounded turbulence research. On the other hand, the OFI measurements are of very high quality. Thus, the flow quality is essential in order to obtain reliable turbulence measurements, especially when it comes to fundamental research.

\revs{
The framework developed in the current study allows to adopt different values and distribution types of the uncertain factors and consequently evaluate the corresponding effects on the responses of HWA and OFI. 
From this perspective, the current tool can be seen as a virtual laboratory facilitating our understanding of the propagation of uncertainties in the actual HWA-OFI experiments.  
Upon availability of more detailed information about the experiments and the measured quantities, further development of the current framework to perform more elaborate uncertainty analysis is feasible. 
A possible extension of this work may include more realistic consideration of the $E_{\rm{Pitot}}$ and $E_{\rm{wire}}$ samples in Figs.~\ref{fig:voltagePitotPdf} and \ref{fig:voltageWirePdf} by using summations of Gaussian distributions, which results in a thorough characterization of the uncertainties in $U^{+}$ and $y^{+}$, together with the log-law parameters. 
Based on what \revf{is} shown in the present work, the uncertainty in $U$ measured by HWA varies with flow velocity magnitude, a fact that must be considered in the construction of likelihood function when estimating the parameters of the log-law in the Bayesian framework.
}
\revCom{
A detailed UQ assessment of large-scale experimental facilities like Superpipe, \cite{zagarola}, or CICLoPE, \cite{talamelli14}, including the analysis of the measured streamwise velocity fluctuations, would also be of great value.
Finally, another important extension of the present work would be to consider, in addition to the random measurement errors, the systematic ones which introduce a bias in the measurements.
}


\appendix

\revCom{
\section{Details of global sensitivity analysis (GSA)}\label{sec:GSA_appendix}
Among different possible techniques for variance-based GSA, Sobol sensitivity indices \cite{sobol} are used in this study. 
The main index, indicating the sensitivity of the response $\CR$ to each parameter $q_i$ is defined as:
\begin{equation}\label{eq:sobolMain}
S_{g_i} = \frac{\var \left(\BE(\CR|q_i)\right)}{\var(\CR)} \, ,
\end{equation}
where $\BE(\CR|q_i)$ is the expected value of $\CR$ conditional on $q_i$. 
In order to take into account the interactions between the parameter $q_i$ and other parameters in addition to the individual effect of $q_i$, the total sensitivity index
\begin{equation}\label{eq:sobolTotal}
S_{gt_i} = 1-\frac{\var \left(\BE(\CR|q_{\sim i})\right)}{\var(\CR)} \, ,
\end{equation}
is used, in which $q_{\sim i}$ is a vector containing all parameters except the $i$-th one. 
Note that $\sum_i S_{g_i}\leq 1$, while such a condition does not exist for the summation of the total sensitivity indices. 
Moreover, if the summation of the main indices is much less than 1 then the interaction effects among parameters are more significant. To compute these indices, sampling approaches such as MC and LHS methods, or spectral expansion methods such as PCE, can be implemented. 
For the latter, analytical relations for the sensitivity indices can be derived as shown in Refs. \cite{sudret,tang}.
}

\section{Likelihood function for error-in-variable inverse problem} \label{sec:EiV_likeli}

To construct the likelihood function for Bayesian inference (\ref{eq:bayes_EiV_formul}), assume there are $N$ sets of observations for $x$ and $y$.
In the most general form, $L(\theta,\theta_y ,X,\theta_x,X_t|Y)$ is a function of unknown model and error parameters $\theta$, $\theta_x$, $\theta_y$ in addition to the independent variables $X_t$.
If for the additive EiV model (\ref{eq:bayes_EiV}) with Bayesian inference expressed by (\ref{eq:bayes_EiV_formul}), a multivariate Gaussian distribution with covariance matrix of the following form:
\begin{equation}\label{eq:covMatrixW}
W_j=
\begin{bmatrix}
	\sigma_{x_j}^2 & \cov(x_j , y_j) \\
	\cov(x_j , y_j ) & \sigma_{y_j}^2
\end{bmatrix} 
\end{equation}
is assumed for describing the errors in the observations made for each of the $N$ datasets, then the block-diagonal $(2N \times 2N)$ covariance matrix $V$ with blocks $W_1,W_2,\cdots, W_{N}$ can be constructed.
As a result, the corresponding likelihood function can be written as:
\begin{equation}\label{eq:likeliMultiGen}
L(\theta,\theta_y ,X,\theta_x,X_t|Y) = \left[ (2\pi)^{2N} \det(V) \right]^{-1/2} \exp \left( -\frac{1}{2} Z V^{-1} Z^T \right) \,,
\end{equation}
in which:
$$
Z=\left[ (x_1-x_{t_1}),(y_1-g(x_{t_1},\theta)) ,\cdots, (x_N-x_{t_N}),(y_N-g(x_{t_N},\theta))  \right] \,.
$$
In the special case of mutually uncorrelated observed data $X$ and $Y$, the above likelihood function is simplified as: 
\begin{eqnarray}
L(\theta,\theta_y ,X,\theta_x,X_t|Y) &=& L(\theta,\theta_y ,X_t|Y) L(\theta_x,X_t|X) \nonumber \\
&=& \left[ (2\pi)^{N} \det(V_y) \right]^{-1/2} \exp \left( -\frac{1}{2} (Y-g(X_{t},\theta)) V_y^{-1} (Y-g(X_{t},\theta))^T \right) \nonumber \\
&\cdot& \left[ (2\pi)^{N} \det(V_x) \right]^{-1/2} \exp \left( -\frac{1}{2} (X-X_t) V_x^{-1} (X-X_t)^T \right) \,. \label{eq:likeliMultiUncor}
\end{eqnarray}
Here, $V_x$ and $V_y$ are $N$ by $N$ diagonal matrices with elements $\sigma_{x_j}^2$ and $\sigma_{y_j}^2$, respectively.

As addressed in Ref. \cite{dellaportas}, in the most general case in addition to $p$ model parameters $\theta$, we need to obtain $N$ actual independent variables $X_t$, and $3N$ error parameters $\sigma_x^2$, $\sigma_y^2$, and $\cov(x,y)$.
In the current study it has been observed that for accurate estimation of error parameters, using informative priors (in particular, in inverse-gamma family) is unavoidable. 
It is noteworthy that if in a laboratory experiment \revs{enough number of samples are taken for each dataset, we can use the associated estimates for the error parameters} and hence reduce the total number of parameters in the inverse problem.


\section*{Acknowledgments}
RV and PS acknowledge the financial support from the Swedish Research Council (VR) and the Knut and Alice Wallenberg Foundation. The authors also thank Profs. Hassan Nagib and David Williams from IIT (Chicago) for making the HWA dataset available for the present study, \revs{and Prof. Per L{\"o}tstedt from Uppsala University for valuable comments and discussions on uncertainty quantification}.

\bibliographystyle{plain}
\bibliography{bibUQuukth}

\end{document}